\documentclass[useAMS,usenatbib]{mn2e}
\usepackage{graphicx}
\usepackage{color}
\usepackage{subfigure}
\newcommand{\Mo}{M_\odot}
\newcommand{\Mwd}{M_{\rm WD}}
\newcommand{\Rwd}{R_{\rm WD}}

\title[New Comprehensive X-ray Spectral Model from the PSAC in IPs]
{A New Comprehensive X-ray Spectral Model from the Post-shock Accretion Column in Intermediate Polars}
\author[T. Hayashi and M.Ishida]{Takayuki Hayashi$^{1, 2}$
\thanks{E-mail:thayashi@astro.isas.jaxa.jp} 
and Manabu Ishida$^{1, 2}$
\\
$^{1}$The Institute of Space and Astronautical Science/JAXA, 3-1-1 Yoshinodai, Chuo-ku, Sagamihara 252-5210\\
$^{2}$Department of Physics, Tokyo Metropolitan University, 1-1 Minami-Osawa, Hachioji, Tokyo 192-0397}
\begin{document}

\date{}

\pagerange{\pageref{firstpage}--\pageref{lastpage}} \pubyear{2012}

\maketitle

\label{firstpage}

\begin{abstract}
We model the post-shock accretion column (PSAC) of 
intermediate polars (IPs) 
with the 
specific accretion rate being floated in the range 
between 0.0001 and 100 g~cm$^{-2}$~s$^{-1}$ and 
the metal abundance in the range between 0.1 and 2 times of the solar, 
and taking into account the gravitational potential with radial dependence,
non-equipartition between ions and electrons, and ionization non-equilibrium. 
We fully take into account the 
dipole 
geometry for the PSAC. 
The specific accretion rate significantly affects the structure of the PSAC, 
and there is a critical rate below which the profiles of the density 
and temperature distributions deviate from those of the standard model. 
This happens when the specific accretion rate is 1 and 
30~g~cm$^{-2}$~s$^{-1}$ for the 0.7 and 1.2~$M_\odot$ white dwarf (WD), respectively, 
or the height of the PSAC becomes
1\% of the white dwarf radius. 
Below the critical specific accretion rate, the present standard model is no longer valid.
We calculate the spectra of the 
PSACs 
with 
the density and temperature distributions described above. Input parameters are the mass of the WD, the specific accretion rate, and the metal abundance. The spectral shape is constant and consistent with that of the standard model if the specific accretion rate is larger than the critical value, except for density-dependent emission lines. Below the critical specific accretion rate, on the other hand, the spectra
soften as the specific accretion rate decreases. 
Associated with this, the maximum temperature of the PSAC becomes significantly lower than that of the standard model below the critical specific accretion rate.
Although the 
ionization non-equilibrium are also considered in 
the spectral calculation,
the effects are limited 
because the radiation from ionization non-equilibrium plasma
is a few percent of  the whole at most.
\end{abstract}


\section{Introduction}
Magnetic cataclysmic variables (mCVs) are binary systems made up of a Roche Lobe-filling late type star
and a magnetized ($B >$ 0.1 MG) white dwarf (WD).
There are two subclasses in the mCV. One 
is the polar
in which the magnetic field of the WD is so strong ($B > 10$~MG) 
that the WD rotation and the binary motion are synchronized. 
The other is the intermediate polar (IP),  the subject of 
this paper,
where the WD rotation is not synchronized with the binary revolution. 
In IPs, matter split over the Roche lobe of the secondary star initially forms 
an accretion disk, is funneled by the strong magnetic field 
within the Alfv$\acute{\rm e}$n radius,
and falls toward the WD surface nearly at the free fall velocity.
Since the accreting matter becomes highly supersonic 
as it descends along the magnetic field line, 
a strong shock is formed and the matter is heated up to a temperature of order 10~keV.
The high temperature plasma 
formed below the shock front is 
cooled via optically thin thermal plasma emission,
and finally settles onto the WD surface.
The plasma flow in 
the downstream side of the shock is 
called the post-shock accretion column (PSAC).
X-ray spectra emitted from the IPs reflect
structure of the PSAC 
such as radial distributions of the temperature and density. 
One of the most important parameter is the maximum temperature ($T_{\rm max}$) of the plasma.
Since this quantity reflects the depth of the gravitational potential
of the WD ($\propto M_{\rm WD}/R_{\rm WD}$),
the WD mass can be estimated from the X-ray spectra 
with the aid of a theoretical WD mass-radius relation,
for example, described by \cite{1972ApJ...175..417N},
\begin{equation}
   R_{\rm WD} = 7.8\times10^8\left[\left(\frac{1.44~\Mo}{\Mwd}\right)^{2/3}
   -\left(\frac{\Mwd}{1.44~\Mo}\right)^{2/3}\right]^{1/2}~{\rm cm}.\label{eq:M_R}
\end{equation}

\cite{1973PThPh..49..776H} first considered 
a steady and spherically symmetric 
accretion flow 
onto WDs.
He discussed that  
near the WD surface a shock is formed and a hot plasma between the shock front
and the WD surface emits thermal radiation.
Although he simply carried out arithmetic operations, 
he estimated some physical quantities, for example, 
an effective temperature and emission measure
of the post-shock plasma.
Shortly after \cite{1973PThPh..49..776H},
\cite{1973PThPh..49.1184A} analytically calculated the distributions 
of the temperature and the density along 
the plasma flow under the assumption
that the thickness of the emission region,
which corresponds to the height of the PSAC
, is negligible 
compared with the WD radius.
This assumption means that the gravitational potential can be considered as 
constant 
throughout 
the PSAC.
Since his assumption appeared to be reasonable for high accretion rate systems 
and the analytically expressed temperature and density distributions
are easily accommodated to the
observed spectra,
Aizu model had been used for 
evaluation of the observed spectra
and WD mass estimation until 
a few tens of years after his publication,
for example, in \cite{1997ApJ...474..774F}.

After that, a lot of theoretical studies were performed for 
the PSAC 
 (e.g.~\citealt{1983ApJ...268..291I}, \citealt{1996A&A...306..232W},
\citealt{2005A&A...440..185C}, \citealt{2005MNRAS.360.1091S} and \citealt{2007MNRAS.379..779S}).
They included two-fluid effects, Compton cooling, effect of gravitational potential or dipolar geometry.
Their calculations are, however, too complex to be applied to observed X-ray spectra.
\cite{1994ApJ...426..664W} and \cite{1998MNRAS.293..222C}, on the other hand, beautifully simplified
the PSAC 
model formulation, which enables us to use it for evaluating the observed spectra, and to
extract information of the PSAC and the WD.
They assumed one-temperature (i.e. equipartition between ions and electrons), 
one-dimensional and cylindrical geometry for the PSAC. 
\cite{1999MNRAS.306..684C} continued along the path of improvements to that technique 
by addressing the elimination of a negligible shock height assumption. 
In so doing, they explored and elucidated clearly for the first time the effects of 
including a radially varying gravitational potential along the PSAC.
The model of \cite{1999MNRAS.306..684C} have been used for the resent WD mass measurements with X-ray spectra
(\citealt{2000MNRAS.316..225R}, \citealt{2005A&A...435..191S}, \citealt{2009A&A...496..121B}, 
\citealt{2010A&A...520A..25Y} and so on) and is said to be the 
present standard model of 
the PSAC. 
In fact, the standard model can reproduce observed spectra well.
While \cite{1999MNRAS.306..684C} considered the radially varying-
gravitational potential,
they hardly discussed the difference of the mass accretion rate per unit area called "specific accretion rate". 
Moreover, some studies using the standard model
(e.g. \citealt{2005A&A...435..191S} and \citealt{2010A&A...520A..25Y} for light WDs)
noted that the PSAC 
structure is 
not influenced so much, and hence
the WD masses estimated with the observed spectra are altered 
by the specific accretion rate only a little.
As a matter of fact, \cite{2010A&A...520A..25Y} 
investigated the influence of the specific accretion rate $a$
on the WD mass estimation in the range $a = $ 0.1--10 
~g~cm$^{-2}$~s$^{-1}$.
As a results, they showed that the estimated WD mass 
is affected 
by 
less than $\sim$ 30\% for a WD less massive than 
1.2 $\Mo$ where most WDs are likely to belong.
However, some observations suggest that the specific accretion rate 
distributes in a wider range. 
For instance, the accretion rate 
of AE Aquarii 
and V1223 Sagittarii are
estimated at $\dot{M} \sim 10^{14}$~g~s$^{-1}$
\citep{1991ApJ...382..290E} 
and 8.4 $\times$10$^{16}$~g~s$^{-1}$ \citep{2011PASJ...63S.739H} with the standard model, respectively.
Furthermore, 
\cite{1998MNRAS.293..222C} obtained the specific accretion rate of 
EX Hydrae 
of 0.001~g~cm$^{-2}$~s$^{-1}$ as the best fit parameter,
which is three orders of magnitude lower than that assumed in the standard model,
although they did not take into account the radially varying-
gravitational potential.   
Although 
these results are indirect evaluations,
they suggest
the specific accretion rate may be different by 
more than a few orders magnitudes among the IPs.

One of the remaining major issues of the standard model is that 
the mass of the WD and the height of PSAC in some IPs evaluated 
by the standard model are inconsistent with that derived from 
observations
in low accretion rate or massive WD systems. 
The maximum temperature of the plasma in the peculiar IP AE Aquarii is 4.6~keV \citep{2006ApJ...639..397I}.
Based on the standard model, the low maximum temperature means that 
the WD 
 of the AE Aquarii is no more massive than 0.2~$\Mo$ (see figure 2 in \citealt{2010A&A...520A..25Y} ),
which is only 
one-fourth of the 0.79~$\Mo$ measured with 
the line Doppler measurement in optical band 
\citep{1996MNRAS.282..182C}.
In this system, owing to its fast WD spin with a period of $\sim$ 33 sec 
\citep{1979ApJ...234..978P},
the propeller effect plays an important roll and blows away most of the accreting matter \citep{1997MNRAS.286..436W},
which leads very low mass accretion rate $\dot{M} \sim 10^{14}$~g~s$^{-1}$.
The WD mass of EX Hydrae, 0.42$\pm0.02$~$\Mo$ \citep{2010A&A...520A..25Y} 
estimated with the standard model 
contradicts 
the 
optical measurement 
0.79$\pm$0.26~$\Mo$ \citep{2008A&A...480..199B}.
We believe that the latter measurement
is fairly reliable because EX Hydrae is a double-lined eclipsing IP.
This system is also a low accretion-system, $\dot{M}$ = 2.8$\times$10$^{15}$~g~s$^{-1}$ \citep{2005A&A...435..191S}.
The WD mass of a nova IP GK Perseus is estimated at 1.15~$\Mo$ 
with "nova universal decline law"  \citep{2007ApJ...662..552H},
which is much greater 
than the X-ray estimations
, 0.59$\pm$0.05~$\Mo$ by \citet{2005A&A...435..191S},
0.90$\pm$0.12~$\Mo$ by \citet{2009A&A...496..121B} 
and $0.92^{+0.39}_{-0.13}$~$\Mo$	by \citet{2009MNRAS.392..630L}.
Although this IP is very high accretion-rate system of 81.5$\times$10$^{16}$~g~s$^{-1}$,
the massive WD \citep{2005A&A...435..191S} 
causes a relatively large error in the mass estimation.
As for the height of the PSAC in EX Hydrae, 
the standard model expects 
about 2$\times10^6$~cm, comparable to 0.2\% of the WD radius \citep{2010A&A...520A..25Y},
\cite{1998MNRAS.295..167A} derived the height 
as tall as the 
WD radius. 
These discrepancies show that the standard model should further be improved 
in some cases, especially for the cases of the lower specific accretion rate and of the massive WD.

In order to resolve the issues described above, 
we modify the standard model in the following four points;
(1) the specific accretion rate, which is fixed at 1 g~cm$^{-2}$~s$^{-1}$ in the standard model,  is 
floated in the wide range from 0.0001 to 100~g~cm$^{-2}$~s$^{-1}$.
The lower specific accretion rate enhances the height of 
the PSAC 
because the density 
becomes lower, 
which results in a 
longer cooling time. 
The PSAC extension further requires; 
(2) the dipole should be considered as 
the PSAC geometry instead of the cylinder used in the standard model,
which reduces the density further 
because of the funnel shape 
of the PSAC.
Due to the density reduction; 
(3) the non-equilibrium between ions and electrons and 
(4) the ionization non-equilibrium should be taken into account
which are not considered 
in the standard model.
These modifications are more important for the IPs holding the massive WD
because the more massive WD is smaller in size and the PSAC height 
becomes more significant 
compared with the WD radius at a higher 
specific accretion rate.

This paper is organized as follows. 
The calculation scheme is described 
in section~\ref{sec:model}.
In section~\ref{sec:calc}, we investigate the PSAC 
structure at various 
specific accretion rate and metal abundance.
In section~\ref{sec:spe}, the spectra emitted from the PSACs 
are shown and their dependence on the WD mass, the specific accretion rate
and the metal abundance are 
discussed.
Finally, we summarize our results in section~5.

\section{Modeling the Post-Shock Accretion Column}\label{sec:model}
We modeled the PSAC according to the guidelines described in the previous section. 
Following the method of \cite{1999MNRAS.306..684C} and \cite{2005A&A...435..191S},
the distributions of the density and temperature were calculated.
\begin{figure}[!h]
\hspace{5mm}
\includegraphics[width=70mm]{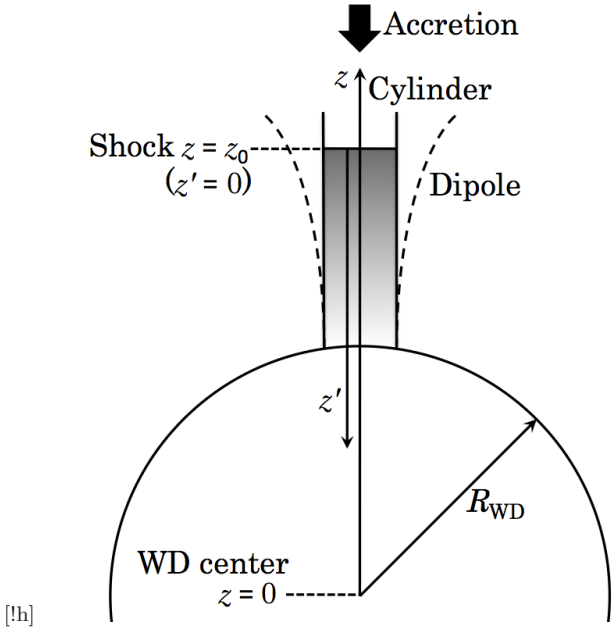}
 \caption{Geometries of the PSAC models.
 Dashed lines show the dipolar geometry.}
 \label{fig:accretion_column}
\end{figure}
The simultaneous differential equation involving the mass continuity equation
\begin{eqnarray}
\frac{\rm{d}}{\rm{d}z}(\rho v S) = 0,\label{eq:conti}
\end{eqnarray}
the momentum equation
\begin{eqnarray}
\frac{{\rm d}}{{\rm d}z}(\rho v^2+P) = -\frac{GM_{\rm WD}}{z^2}\rho-\frac{\rho v^2}{S} \frac{{\rm d}S}{{\rm d} z},\label{eq:momentum}
\end{eqnarray}
the energy equation
\begin{eqnarray}
v\frac{{\rm d}P}{{\rm d}z}+\gamma P\frac{{\rm d}v}{{\rm d}z}=-(\gamma-1)\left(\varepsilon-\frac{\rho v^3}{2S} \frac{{\rm d}S}{{\rm d} z}\right)\label{eq:energy}
\end{eqnarray}
and the ideal-gas law
\begin{eqnarray}
P=\frac{\rho kT}{\mu m_{\rm H}}
\end{eqnarray}
describes the PSAC 
structure. 
Here, $z$ is the spatial coordinate shown in figure \ref{fig:accretion_column} whose origin is the WD center,
$v$ is the bulk velocity, $\rho$ is the mass density, $T$ is the averaged temperature,
$P$ is the thermal pressure of the plasma,
$\gamma = 5/3$ is the adiabatic index, and $\mu = 0.62$ is the mean molecular weight.
$S$ is the cross-section of the PSAC and
assumed to be proportional to a power law function of $z$,
\begin{eqnarray}
S \propto z^{n}.
\end{eqnarray}
The power-law index $n$ of 0 and 3 correspond to cylinder and dipole, respectively.
$\varepsilon$ is the cooling rate via optically thin thermal radiation given by
\begin{eqnarray}
\varepsilon=\left(\frac{\rho}{\mu m_{\rm H}}\right)^2\Lambda(T),
\end{eqnarray}
where $\Lambda$ is the cooling function.
We adopt 
the Collisional Ionization Equilibrium (CIE) cooling function calculated 
by SPEX package \citep{2009A&A...508..751S}.
One may doubt that 
we should use a cooling function that reflects 
the non-equilibrium effects 
correctly.
However, 
the part of the PSAC where the equipartition between the electron and the ion is not achieved 
is of lower density 
by a few orders of magnitude than that in the equilibrium part,
and hence we can use the CIE cooling function throughout the PSAC.
In fact, 
we substituted 
the emissivity of thermal 
bremsstrahlung for the CIE cooling function
in the part where the electron temperature is below 90, 95, 99 and 99.9\% of the averaged temperature.
As a results, all these hybrid cooling functions gave nearly 
identical results.
The second term of the right-hand side of the equation \ref{eq:momentum} 
describes the conversion of thermal energy into the kinematic energy in the PSAC
because of decreasing cross-section as the Venturi nozzle.

The integral of equation (\ref{eq:conti}) is 
\begin{eqnarray}
\rho v S = aS = a (4\pi R_{\rm WD}^2 f) = \dot{M},\label{eq:conti_integral}
\end{eqnarray}
where $a$ is the mass accretion rate per unit area or 
"specific accretion rate" at the WD surface.
$f$ and $\dot{M}$ 
are fractional accretion area 
and the mass accretion rate. 
In the standard model, $a =$ 
1~g~cm$^{-2}$~s$^{-1}$ and $n = $ 0 for any IPs.

Equations (\ref{eq:momentum}) and (\ref{eq:energy}) 
can be transformed 
with $a$ and $z'$ = $z_{\rm 0} - z$,
where $z_{\rm 0}$ is the shock coordinate (see figure \ref{fig:accretion_column}),
\begin{eqnarray}
\frac{{\rm d}v}{{\rm d}z'}&=&g(z')\frac{1}{v}-\frac{1}{a}\frac{{\rm d}P}{{\rm d}z'},\label{eq:momentum_edit}
\end{eqnarray}
\begin{eqnarray}
\frac{{\rm d}P}{{\rm d}z'}=\frac{(\gamma-1)(\varepsilon-\frac{\rho v^3}{2S}\frac{{\rm d}S}{{\rm d}z'})a+g(z')\gamma P\rho}{\gamma P-av},\label{eq:energy_edit}
\end{eqnarray}
where 
\begin{eqnarray}
g(z')&=&\frac{GM_{\rm WD}}{(z_0-z')^2}.\label{eq:gravity}
\end{eqnarray}
Equations (\ref{eq:momentum_edit}) and (\ref{eq:energy_edit}) were solved 
from the top of the PSAC, 
$z = z_0 (z'=0)$ 
to WD surface, $z = R_{\rm WD} (z'=z_{\rm 0}-R_{\rm WD})$
with the following boundary conditions at $z = z_{\rm 0}$
assuming the strong shock at the top of the PSAC,
\begin{eqnarray}
v_0 &=& 0.25\sqrt{2GM_{\rm WD}/z_0}\\
\rho_0 &=& \frac{a}{v_0},\\
P_0 &=& 3av_0,\\
T_0 &=& 3\frac{\mu m_{\rm H}}{k}v_0^2
\end{eqnarray}
and soft landing,
\begin{eqnarray}
v_{\rm WD} &=& 0~{\rm (at~WD~surface)}.
\end{eqnarray}
The boundary conditions are uniquely given by specifying $M_{\rm WD}$, $a$, and $z_0$. Of them,
the 
shock position $z_0$ matching 
the boundary conditions
was found by iteration (shooting method).
In so doing, we utilize the mass-radius relation of WDs given by equation \ref{eq:M_R}.

Unlike the standard model, we consider 
non-equipartition between ions and electrons.
After ions are immediately heated up by the strong shock,
electrons are heated by Coulomb scattering with the ions.
Time variation 
of the electron temperature is written as 
\begin{eqnarray}
\frac{{\rm d}T_{\rm e}}{{\rm d}t} = \frac{T_{\rm i}-T_{\rm e}}{t_{\rm eq}},
\label{eq:equilibrium}
\end{eqnarray}
where $t_{\rm eq}$ is the time scale of equipartition calculated with
\begin{eqnarray}
t_{\rm eq}  &=& 5.87\frac{A_{\rm e}A_{\rm i}}{n_{\rm i}~\rm cm^{-3}~Z_{\rm e}^2Z_{\rm i}^2\ln\Lambda}\left(\frac{T_{\rm i}~\rm K}{A_{\rm i}}+\frac{T_{\rm e}~\rm K}{A_{\rm e}}\right)^{3/2} {\rm s}.
\label{eq:eq_time}
\end{eqnarray}
\citep{1962pfig.book.....S}.
Here, $T$~(in Kelvin), $A$ and $Z$ are temperature, atomic weight and charge,
and subscript e and i imply 
electron and ion, respectively. $n_{\rm i}$ is the number density of the ion in a unit of cm$^{-3}$.
For practical calculation, we followed the method of \cite{2009ApJ...707.1141W}
in which 
eq.(\ref{eq:equilibrium}) can be rewritten as
\begin{eqnarray}
\frac{{\rm d}\tau}{{\rm d}t} = \frac{2\ln\Lambda}{503}\left<\frac{Z_{\rm i}^2}{A_{\rm i}}\right>\frac{n}{T^{3/2}}\tau^{-3/2}(1-\tau)~{\rm s}^{-1}\label{eq_wong}
\end{eqnarray}
%
where $T$ is the averaged temperature; 
\begin{eqnarray}
T = \frac{n_{\rm e}T_{\rm e}+n_{\rm i}T_{\rm i}}{n_{\rm e}+n_{\rm i}}
\end{eqnarray}
and $\tau$ is relative electron temperature,
\begin{eqnarray}
\tau \equiv \frac{T_{\rm e}}{T}.
\end{eqnarray}
%
$\ln \Lambda$ is the Coulomb logarithm and can be approximated as
\begin{eqnarray}
\ln\Lambda \sim 15.9 + \ln\left(\frac{T_{\rm e}}{10^8~{\rm K}}\right)-\ln\left(\frac{n_{\rm e}}{10^{16}~{\rm cm^{-3}}}\right)^{1/2}.
\end{eqnarray}
The ionization 
of heavy element proceeds 
by 
impacts of electrons heated up by the interaction with the ions.
Therefore, the ionization temperature trails the electron temperature.
The angle bracket term is the mean value 
of the ratio of the square of 
ion charge and the atomic number, which equals 1 for our model of a pure hydrogen and helium gas.
We set $A_{\rm e} = 1/1836$.

\section{Calculation Results}\label{sec:calc}
\subsection{Consistency with Previous Works}
\begin{figure}[!h]
\includegraphics[width=80mm]{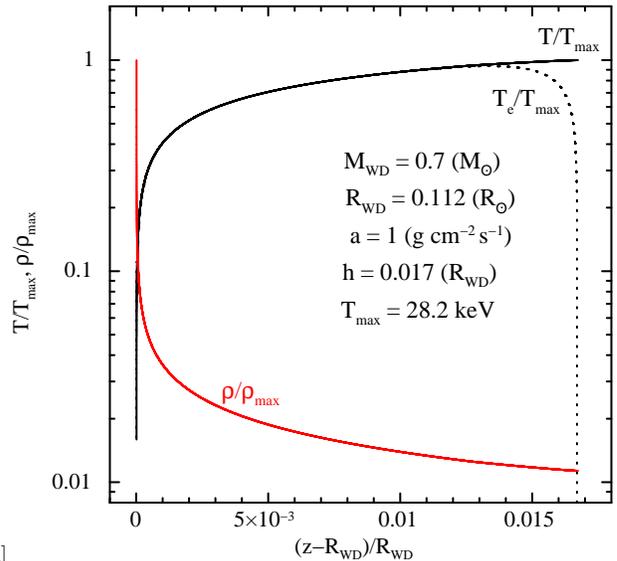}
 \caption{Averaged temperature (black solid line), electron temperature (black dotted line) 
 and density (red solid line) distributions of a cylindrical PSAC 
 for an IP of $\Mo = 0.7~M_\odot$ and $a$ = 1~g~cm$^{-2}$~s$^{-1}$. 
 The parameters are common to figure~2 of \citet{2005A&A...435..191S} and 
figure~3 of \citet{2010A&A...520A..25Y}. }
 \label{fig:norm_T_P}
\end{figure}

Figure \ref{fig:norm_T_P} shows the temperature and density distributions
of the cylindrical PSAC in the case of $a = 1$ g~cm$^{-2}$~s$^{-1}$, $M = 0.7~M_\odot$, and $Z = Z_\odot$. The temperature is
normalized by their 
maximum values. 
The density, on the other hand, is 
normalized at a point one thousandth of the PSAC height, since it diverges at 
the WD surface. 
The profiles of these 
distributions closely resemble those from the 
previous studies of \cite{2005A&A...435..191S} and \cite{2010A&A...520A..25Y}.
Moreover, the shock height of our calculation is 0.017~$R_{\rm WD}$,
which is close to that of \cite{2005A&A...435..191S}, 0.018~$R_{\rm WD}$ and 
\cite{2010A&A...520A..25Y}, 0.013~$R_{\rm WD}$.
The electron temperature does not catch up the averaged temperature 
in the top of 20\% of the PSAC. 
However, 
the density of that region is low. 
This suggests
that the effect of non-equipartition between ions and electrons on 
the X-ray spectrum is limited. 

\subsection{Dependency on the specific accretion rate}
We investigated influence of 
the specific accretion rate, which was fixed at $a = 1$ g~cm$^{-2}$s$^{-1}$ in the standard model, 
on the PSAC 
structure in the range 
between 0.0001 and 100~g~cm$^{-2}$~s$^{-1}$.

\subsubsection{Density distribution and critical specific accretion rate}
\begin{figure}[!h]
\includegraphics[width=80mm]{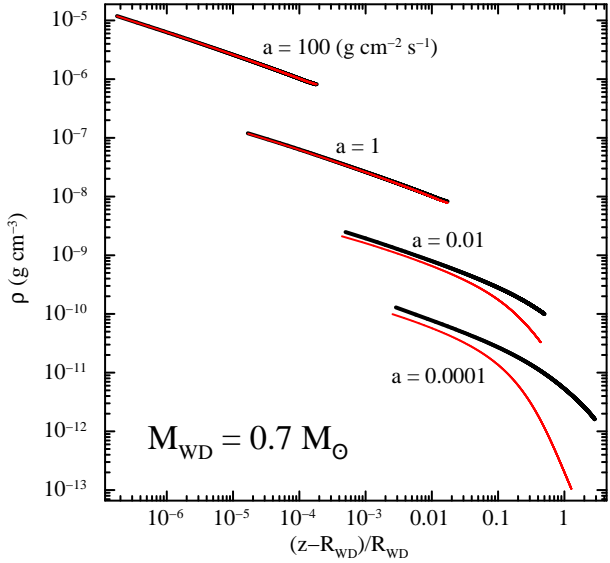}
 \caption{Density distributions of the cylindrical (black) and dipolar (red) PSACs
 for the WD mass of 0.7~$\Mo$ and $a$ of 0.0001, 0.01, 1, 100~g~cm$^{-2}$~s$^{-1}$.
The right ends of the distributions correspond to the tops of the PSACs.
The other ends are terminated at 0.1\% 
of the PSAC height.}
 \label{fig:density_geocomp}
\end{figure}

Figure \ref{fig:density_geocomp} shows the density distributions 
of 
the cylindrical and dipolar PSAC with the specific accretion rate 
in the case of $M_{\rm WD} = 0.7~M_\odot$
of 0.0001, 0.01, 1 and 100~g~cm$^{-2}$~s$^{-1}$.
In this figure, the right ends of each profile corresponds to the shock front.
The other ends are terminated at 0.1\% of a PSAC height of each case.
This figure means that the PSAC becomes taller 
with a lower specific accretion rate due to a longer cooling time.
When the specific accretion rate is sufficiently high 
($a \ga$ 
1
g~cm$^{-2}$~s$^{-1}$ for the 0.7~$\Mo$ WD), 
the density increases toward the WD surface with a power-law function
of the distance from the WD surface, and its profile agrees between the dipolar and cylindrical
geometries.
On the other hand, when the specific accretion rate is 
sufficiently low ($a \ll 1$ g~cm$^{-2}$s$^{-1}$ for the 0.7$M_\odot$ WD)
the density distribution deviates 
from the power law. 
At the same time, 
the density distributions of the dipolar PSACs
become different from those 
of the cylindrical
because the difference between the two geometries emerges
when the PSAC extends upwards and becomes compatible 
with the WD radius.
We hereafter refer to the specific accretion rate below which the density profile of the cylindrical PSAC starts to deviate from that of the dipolar PSAC as the critical specific accretion rate $a_{\rm crit}$. This definition of $a_{\rm crit}$ implies that the standard model is no longer valid in the regime $a < a_{\rm crit}$. We systematically investigated $a_{\rm crit}$ as a function of the WD mass and found that $a = a_{\rm crit}$ occurs when the PSAC height $\simeq 0.01R_{\rm WD}$, irrespective of the WD mass; for instance it is 1 and 30 g~cm$^{-2}$s$^{-1}$ for the 0.7 and 1.2$M_\odot$ WD, respectively. We refer to the readers to consult figure.~5 to find $a_{\rm crit}$ for a WD with any given mass.

\subsubsection{Temperature distribution}

\begin{figure*}[!h]
\begin{center}
\begin{tabular}{cc}
\includegraphics[width=80mm]{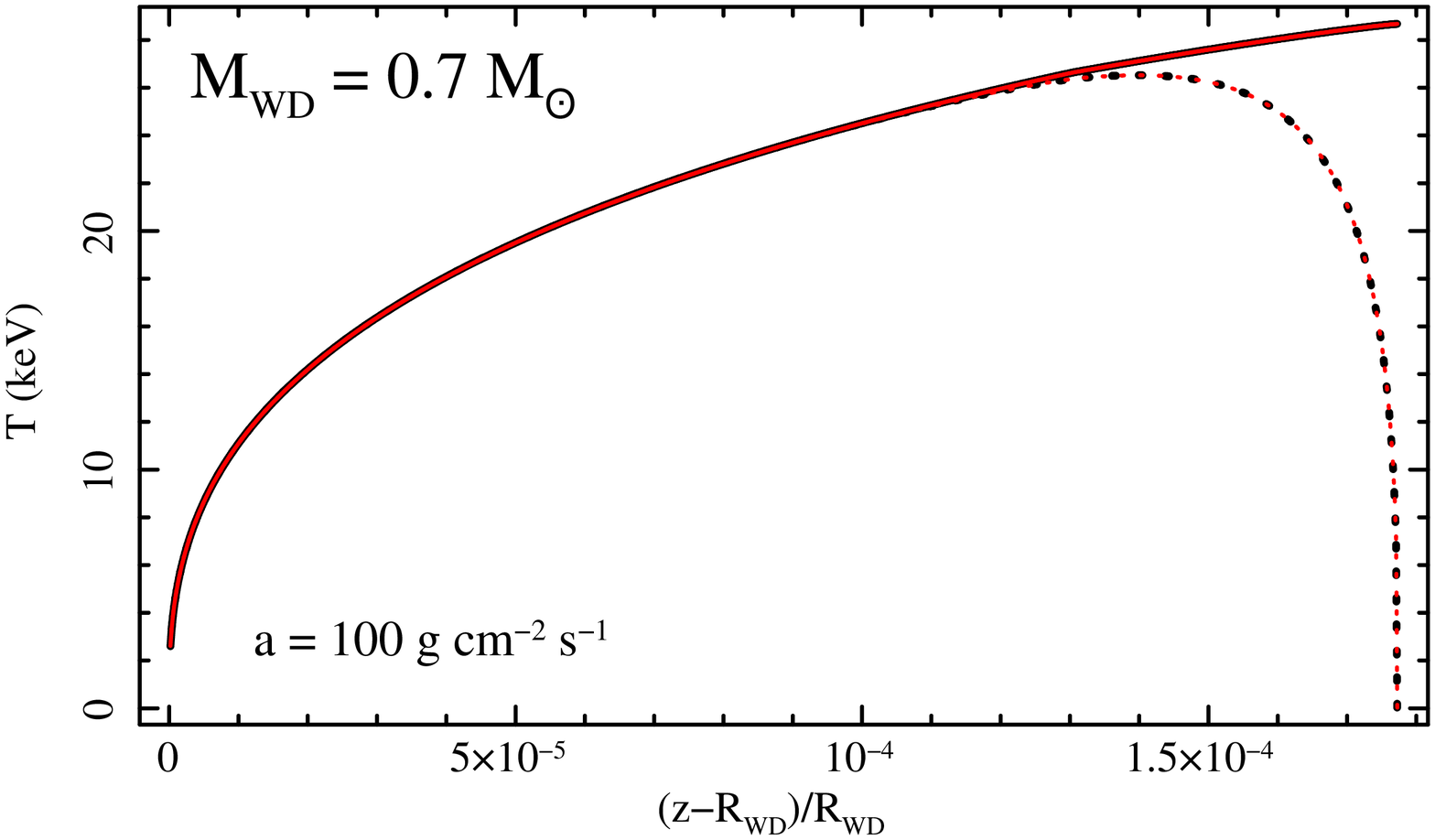}&
\includegraphics[width=80mm]{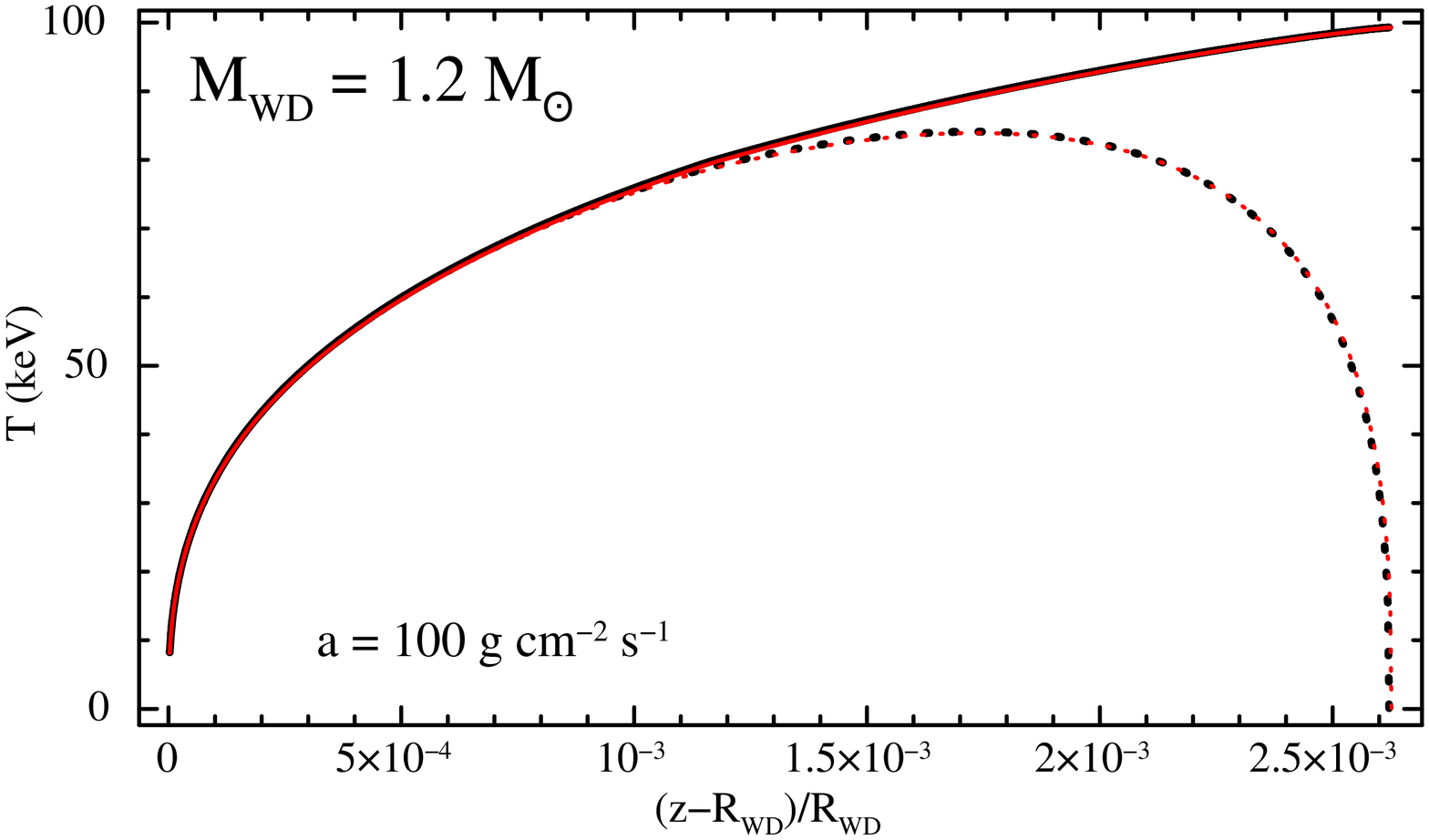}\\
\includegraphics[width=80mm]{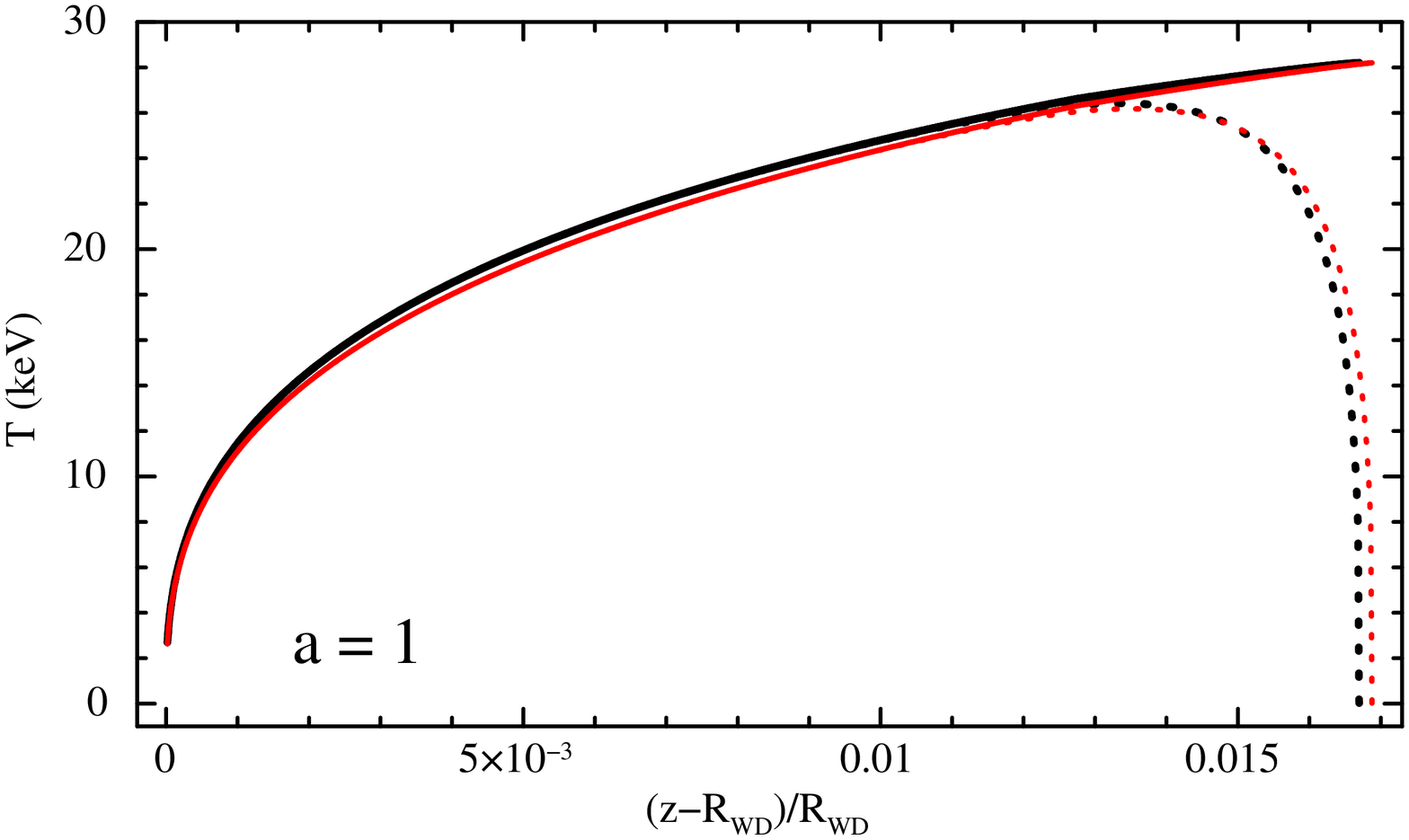}&
\includegraphics[width=80mm]{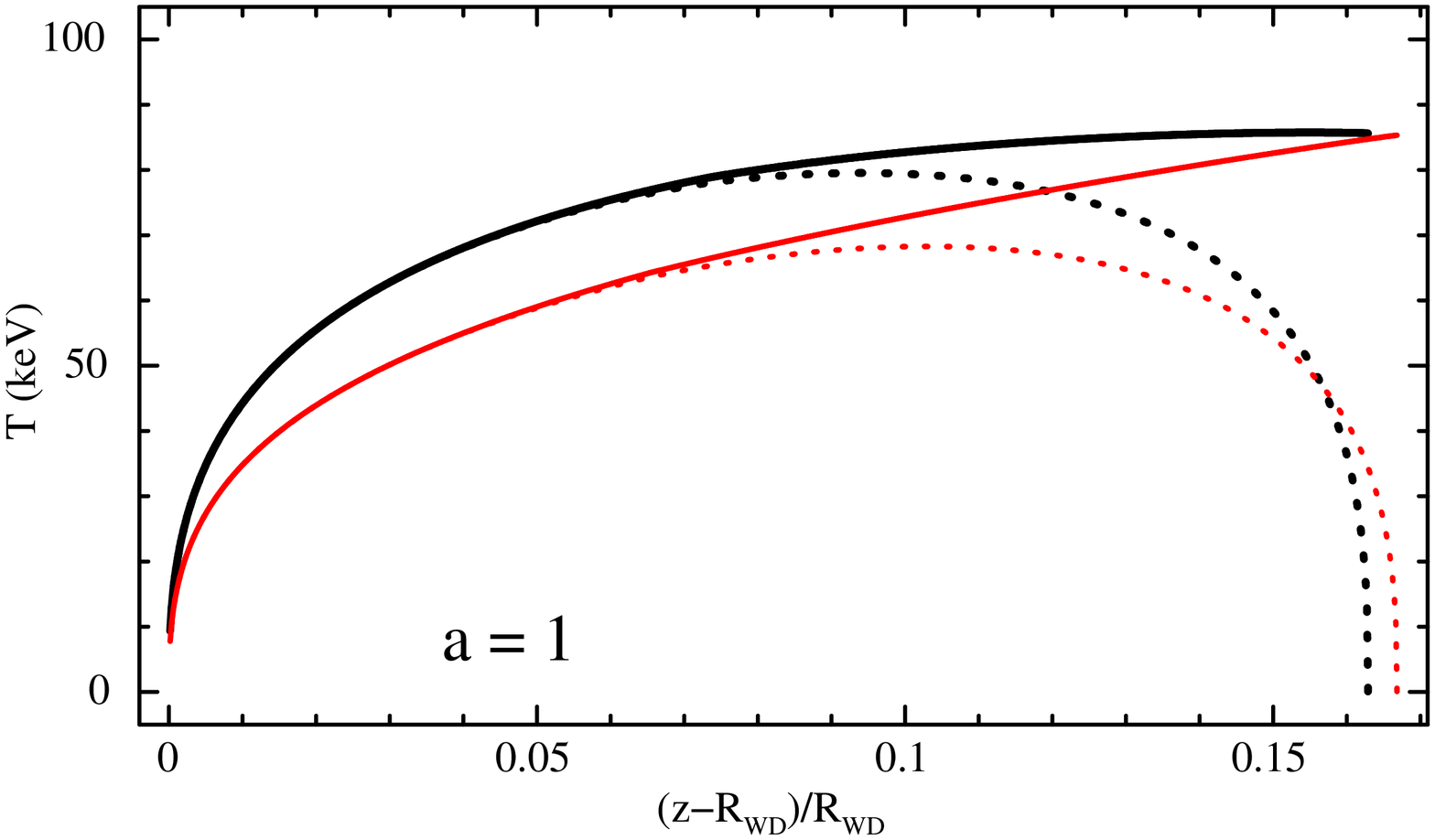}\\
\includegraphics[width=80mm]{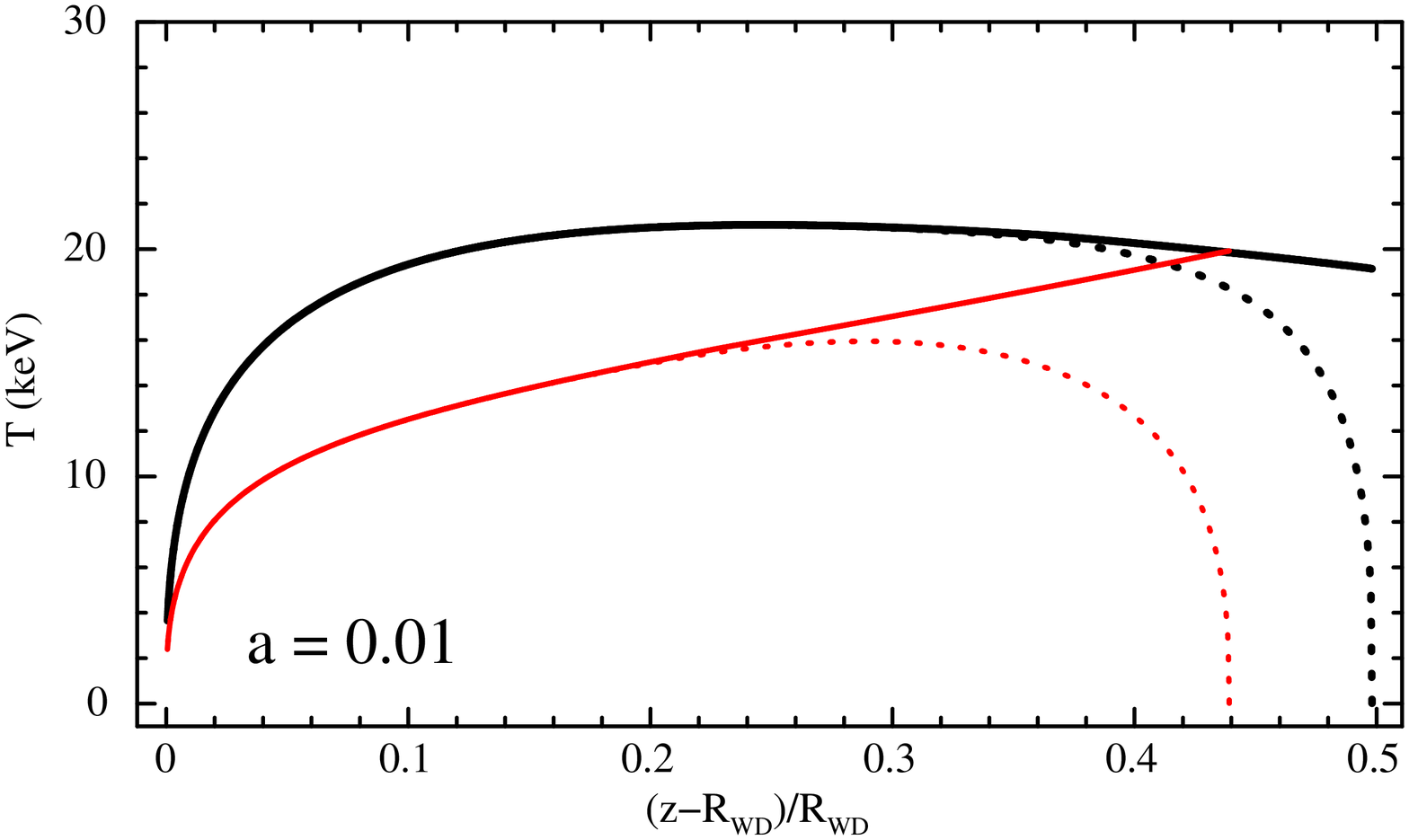}&
\includegraphics[width=80mm]{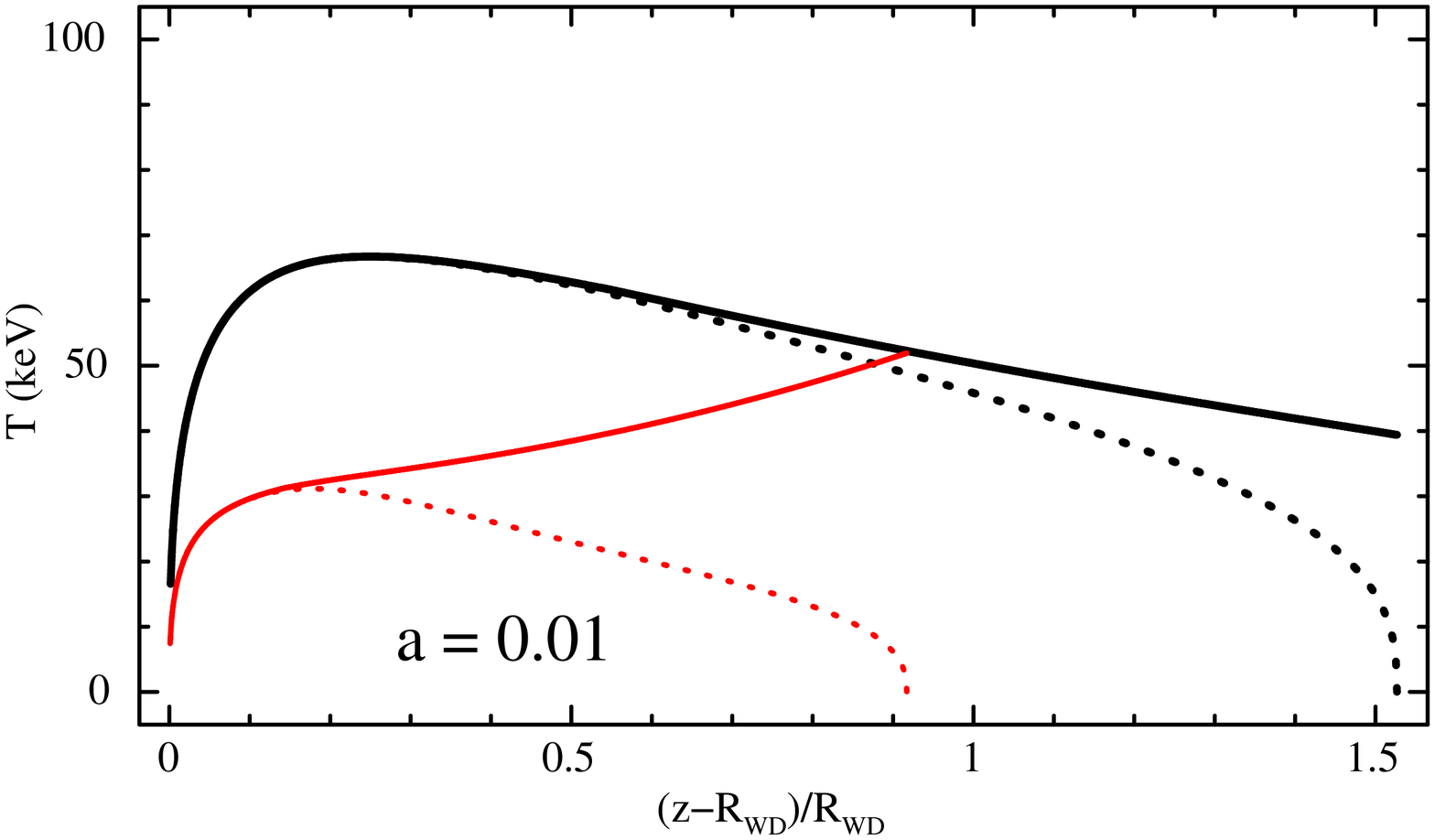}\\
\includegraphics[width=80mm]{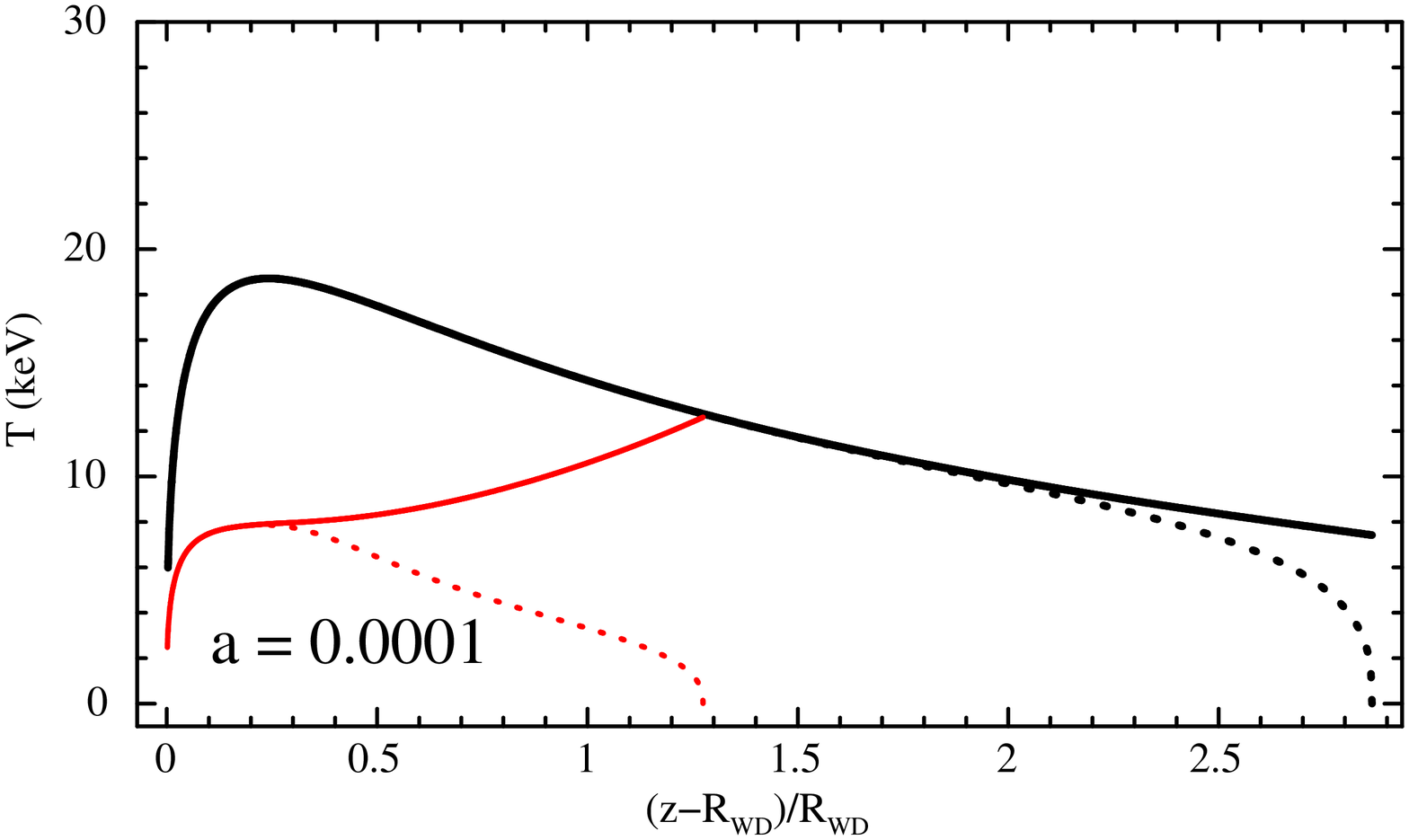}&
\includegraphics[width=80mm]{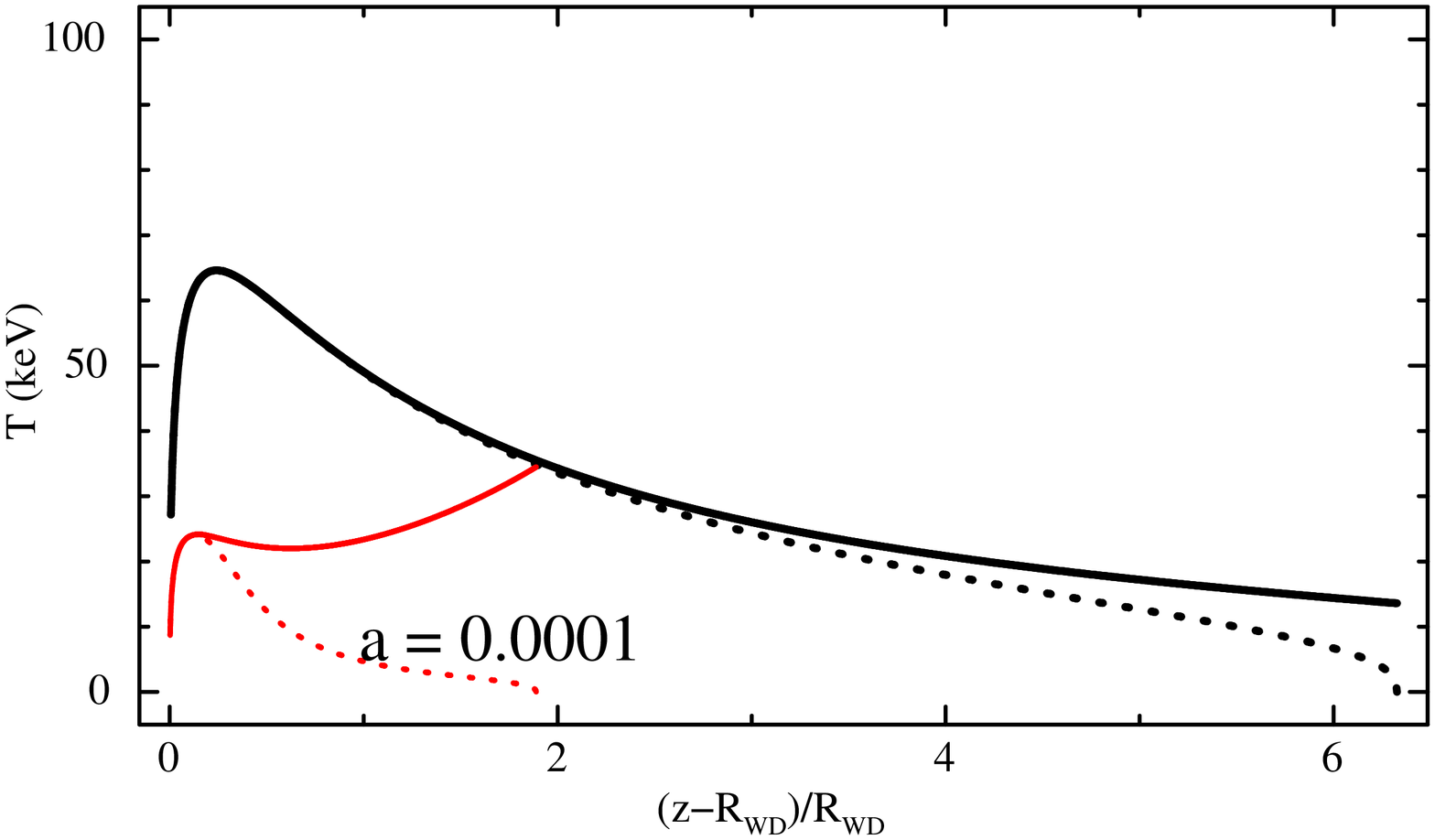}\\
\end{tabular}
\end{center}
 \caption{Averaged (solid) and electron (dotted) temperature distributions 
 of the cylindrical (black) and dipolar (red) PSACs
 for the WD mass of 0.7 (left columns) and 1.2 (right columns)~$\Mo$ and 
 $a$ of 0.0001, 0.01, 1, 100~g~cm$^{-2}$~s$^{-1}$ from bottom to top panels.}
 \label{fig:temperature_geocomp}
\end{figure*}

The distributions of the averaged and electron temperature 
are shown in figure \ref{fig:temperature_geocomp}.
Like 
the density distributions, the shapes of the temperature distributions
agree between
the 
geometries 
if $a \ga a_{\rm crit}$. 
If, on the other hand, 
$a \ll a_{\rm crit}$, 
the averaged temperature reduces from that of the standard model over the whole PSAC
and its distribution flattens.
Although the averaged temperature 
monotonically 
decreases toward the WD surface in the standard model, 
that of our cylindrical PSAC, drawn with black color in figure 4, shows a peak in the middle of the PSAC at 
low enough specific accretion rate. 
This is because energy input by gravity overcomes cooling energy loss 
since the low density 
reduces the cooling rate, 
and the tall 
PSAC 
retains larger amount of gravitational energy to be released below the shock front.
For the dipolar cases, on the other hand,
the temperature decrease is even faster than the cylindrical.
This happens whenever a cross section of a subsonic flow shrinks along the streamline, such as the 
Venturi nozzle. 
In such a case, the bulk velocity of the flow increases at the expense of the thermal energy, which reduces the temperature.
The averaged temperature of the dipolar PSAC monotonically 
decreases as the flow descends the PSAC
for the 0.7~$\Mo$ 
WD throughout the range $a =$
0.0001 -- 
100~g~cm$^{-2}$~s$^{-1}$ unlike the cylindrical case.
However, 
that of the 1.2~$\Mo$ WD
shows a local minimum in the middle of the PSAC, 
as shown in the bottom right panel 
of figure \ref{fig:temperature_geocomp}.
The heat 
transfer from the ion to the electron becomes slower 
with a lower density. 
At a very low specific accretion rate such as 0.0001~g~cm$^{-2}$~s$^{-1}$
the non-equipartition area extends 
over the 
80\% region of the PSAC.
%

\subsubsection{PSAC height}
\begin{figure}[!h]
\includegraphics[width=80mm]{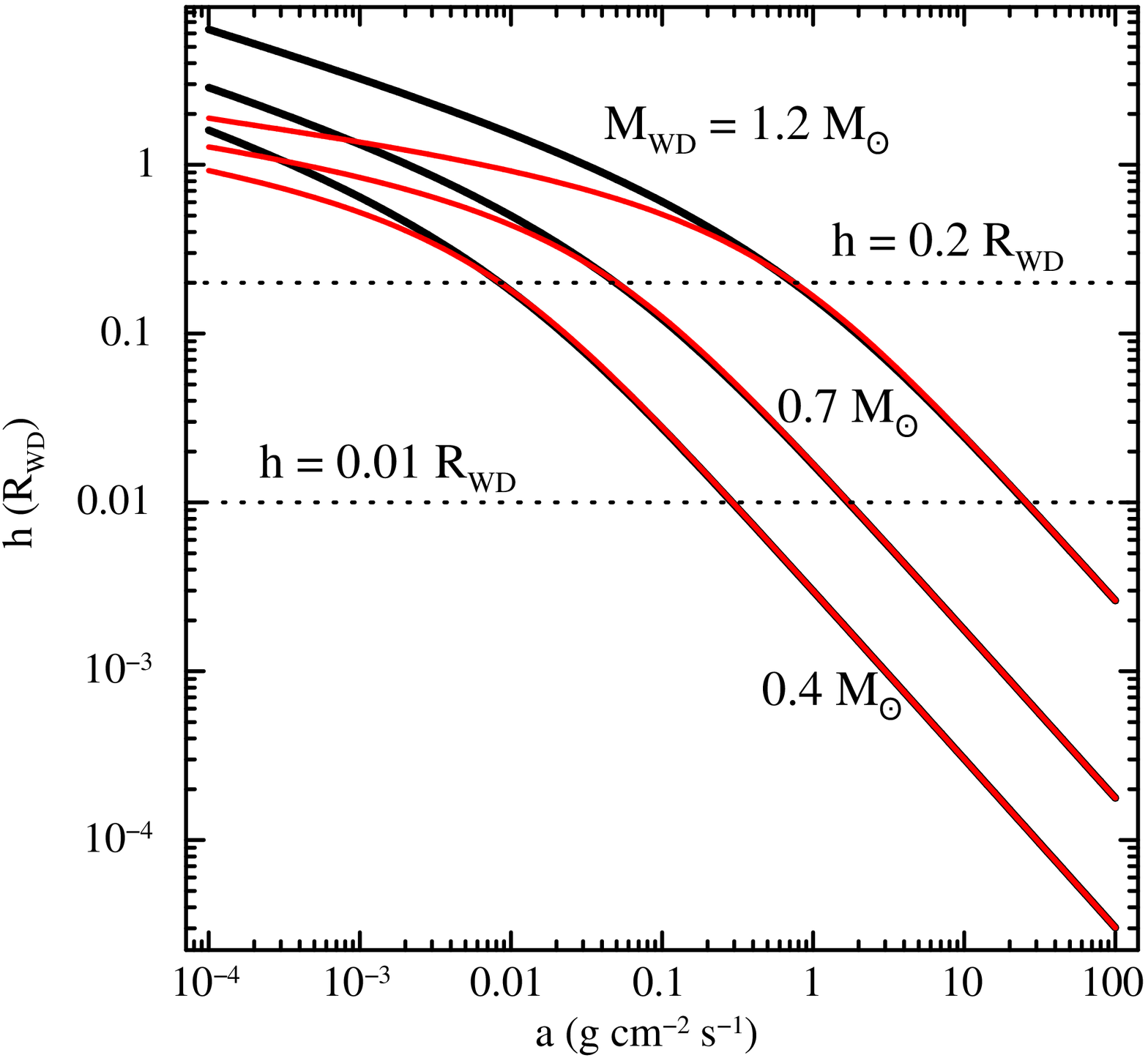}
 \caption{The PSAC heights for the 0.4, 0.7 and 1.2 $\Mo$ 
 WDs as a function of the specific accretion rate.
 Black and red lines show the cylindrical and dipolar cases, respectively.
 The horizontal dotted line represents 20\% of the WD radius, 
 which defines the threshold specific accretion rate.
 Note that the heights in this figure are normalized by each WD radii.}
 \label{fig:a-h}
\end{figure}

Relations between the height of the PSAC and the specific accretion rate 
are shown in 
figure \ref{fig:a-h} for the 0.4, 0.7 and 1.2 $\Mo$ 
WDs.
The PSAC constantly extends upwards with the lower specific accretion rate,
but the slope of the PSAC height abruptly changes at a certain 
value of $a$. 
At around the high end of $a$,
the height is proportional to 
$a^{-1}$,
and
the height of the PSAC 
is almost identical between 
the two, 
PSAC geometries,
whereas at around the low end of $a$
the heights are in proportion to $a^{-0.3}$ and $a^{-0.15}$ 
for the cylindrical and dipolar PSACs, respectively. 
It is interesting to note that the transition between these two regimes occurs at the PSAC
height 
of
about 0.2~$\Rwd$ for 
any mass of the WDs. 
The 
transition appears in larger specific accretion rate 
for a more massive WD, 
because the PSAC height easily becomes 
significant relative to 
WD radius due to its small radius.
The 
radial extent of the PSAC 
can be as large as the WD radius in 
the lower specific accretion rate, which 
may resolve the problem about the observed PSAC height referred section~1.

\subsubsection{Maximum temperature}
\begin{figure}[!h]
\includegraphics[width=80mm]{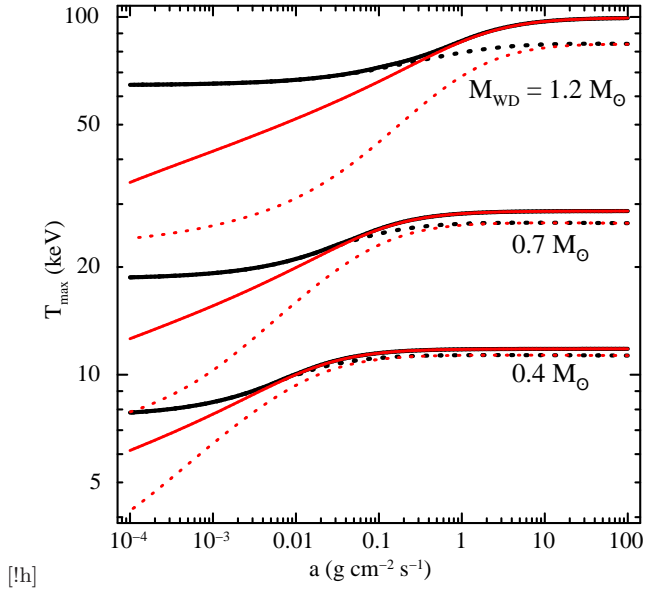}
 \caption{Averaged (solid) and electron (red) maximum temperatures 
relating the specific accretion rate for 0.4, 0.7 and 1.2~$\Mo$ of the WDs.
Black and red lines show the cylindrical and dipolar PSACs, respectively.}
 \label{fig:a-Tmax}
\end{figure}

The maximum 
temperature of electrons and that averaged over ions and electrons
are shown as a function of the specific accretion rate in figure \ref{fig:a-Tmax}
for the 0.4, 0.7 and 1.2 $\Mo$ 
WDs.
The maximum temperature is indicative of the mass of the WD.
In the 
region $a \ga a_{\rm crit}$,
the maximum temperatures of the two geometries 
are 
identical and constant, as is predicted by the standard model.
Below the 
$a_{\rm crit}$,
on the other hand,
the maximum temperatures differ between the two geometries.
In the 
dipolar geometry, the decrease of 
the maximum averaged temperature 
associated with
the decrease of the specific accretion rate is faster 
than that in 
the cylindrical case.
This is because the averaged temperature monotonically decrease for the dipolar PSAC
while the hottest region 
emerges in the middle of the PSAC in 
the cylindrical  geometry, as shown in the bottom four 
panels of figure \ref{fig:temperature_geocomp}.
The electron maximum temperature reduces also faster for the dipolar geometry
because 
the density is smaller than in the cylindrical geometry especially at around the top of the PSAC,
which delays accomplishment of the equipartition.
%

\subsection{Dependency on the metal abundance}
\begin{figure}[!h]
\includegraphics[width=80mm]{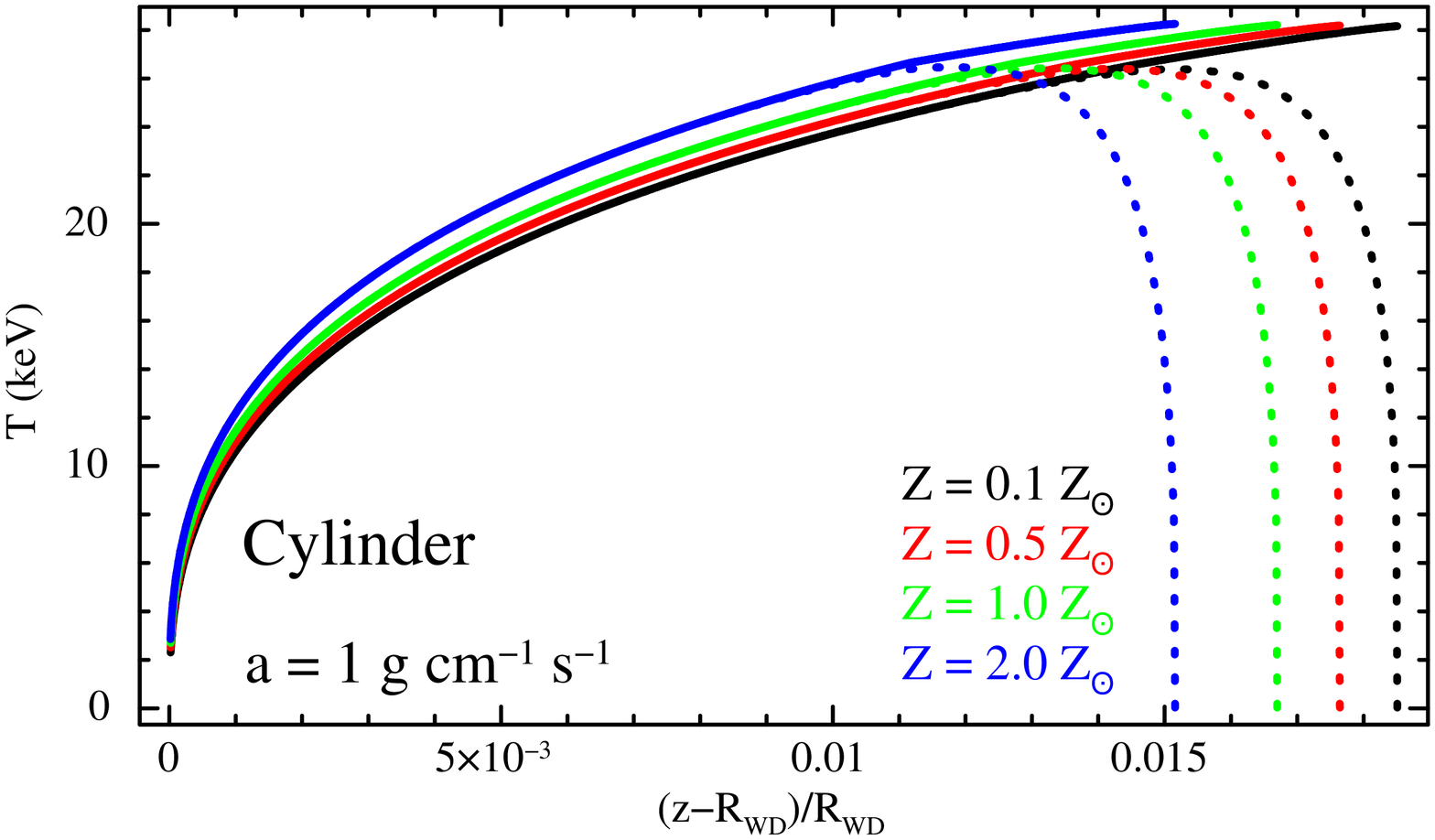}
\includegraphics[width=80mm]{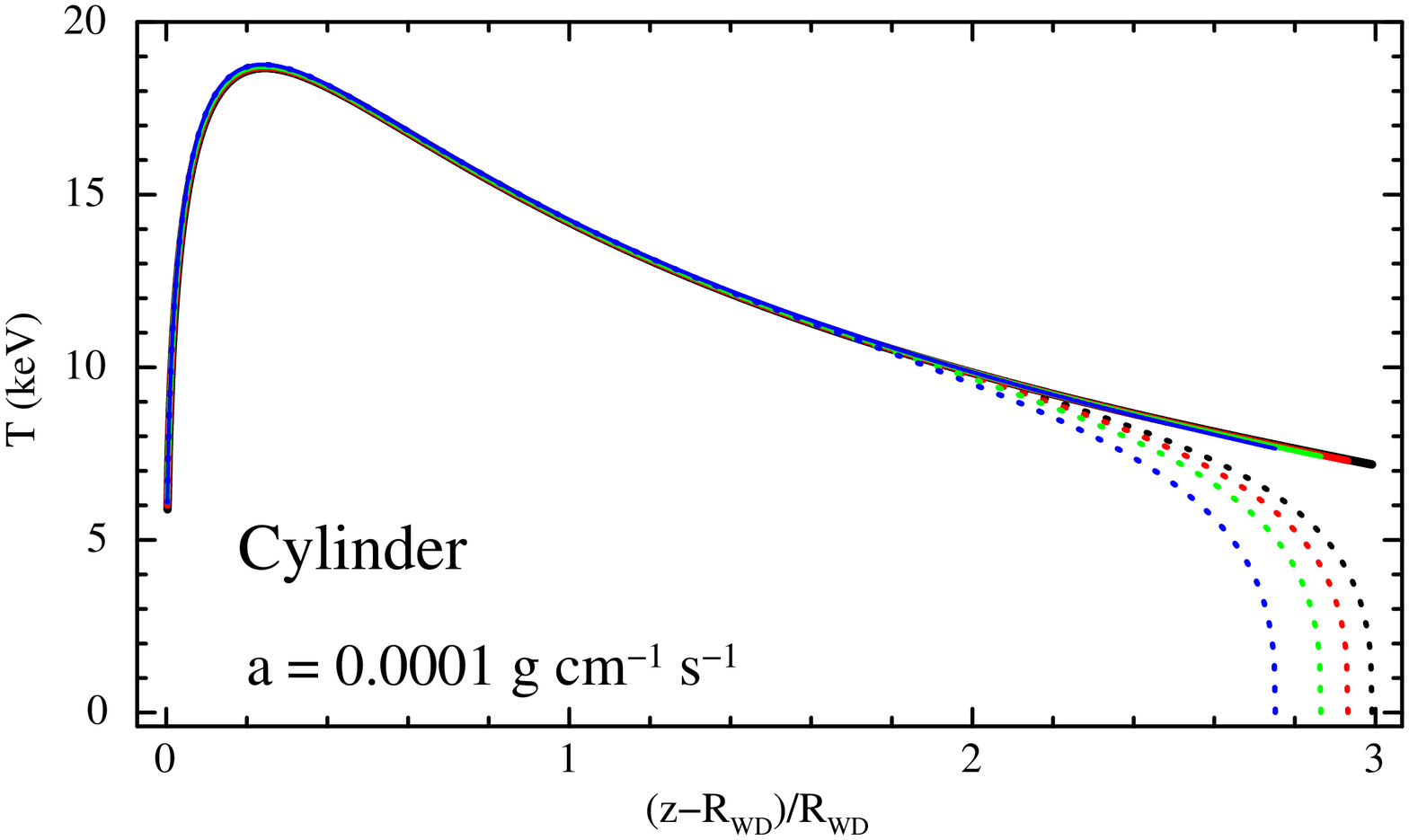}
\includegraphics[width=80mm]{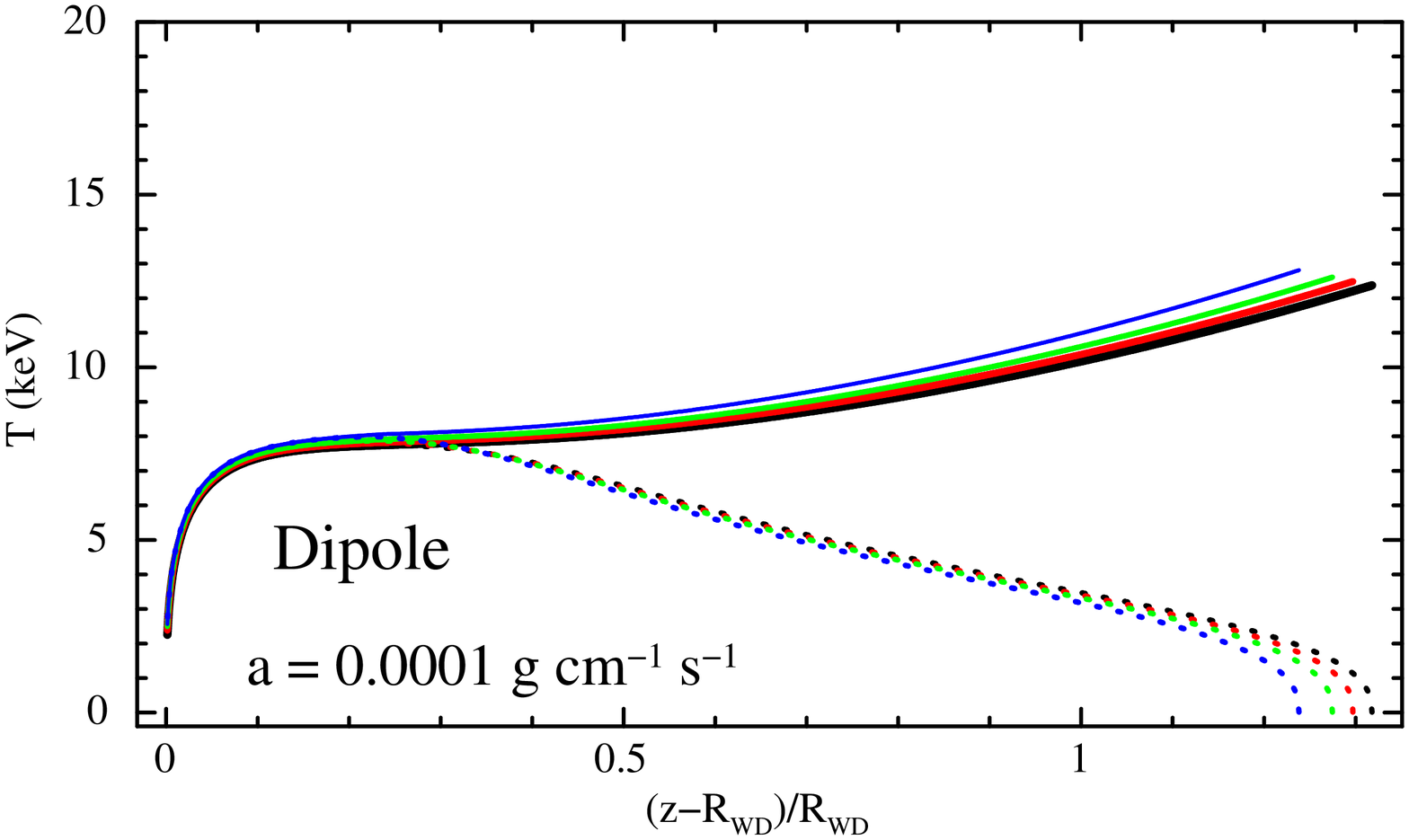}
 \caption{Averaged (solid) and electron (dotted) temperature distributions
  for abundances of 0.1 (black), 0.5 (red), 1.0 (green) and 2.0 (blue) times the 
  solar abundance.
  Top panel shows that the distributions for the cylindrical PSAC 
  and $a$ = 1~g~cm$^{-2}$~s$^{-1}$, which is almost identical to that of the dipolar PSACs
  with the same specific accretion rates. The middle and bottom panels show 
  the distributions for the cylindrical and dipolar PSACs 
  with $a$ = 0.0001~g~cm$^{-2}$~s$^{-1}$. The WD mass of 0.7$M_\odot$ is adopted.}
   \label{fig:abund}
\end{figure}

We investigated influence of metal abundance on the PSAC structure 
as well as the specific accretion rate and the WD mass.
In figure~\ref{fig:abund} we show 
the PSAC structures by assuming 0.1, 0.5, 1.0 and 2.0 $Z_\odot$
with the cylindrical or dipolar geometries.
The top panel of figure~\ref{fig:abund} 
displays examples of 
the cylinder 
PSACs on the 0.7~$M_\odot$ WD. Note that, although cylindrical geometry is assumed, the profiles can be regarded as those of the dipolar PSACs, 
because the assumed specific accretion rate is sufficiently high.
The middle and bottom panels show the temperature distributions
for cylindrical and dipolar 
PSACs on the 0.7~$M_\odot$ WD, respectively.

These results mean that the influence of the metal abundance on the PSAC structure
is even less significant than that of the specific accretion rate.
Moreover, since the abundances of IPs are generally 
in the range from 0.1 to 0.6 times of solar abundance
\citep{2010A&A...520A..25Y}, the influence of the metal abundance 
on the PSAC structure is limited. 
However, the metal abundance significantly affects X-ray spectrum, 
especially line spectra (see section \ref{sec:spe}).
Therefore, we remain 
the metal abundance as an input parameter in the following section.

\section{X-ray spectra}\label{sec:spe}

In this section, we calculate X-ray spectra emitted from the PSACs
using the temperature and density distributions calculated in section \ref{sec:calc}.

\subsection{Spectral calculation method}

In order to calculate the spectra, 
the PSAC is 
divided into evenly spaced one hundred segments within each of
which the 
physical quantities can be regarded as 
constant, 
and then the one hundred partial spectra are summed up.
Spectra emitted from each segments are calculated with one temperature plasma emission models
in SPEX package \citep{1996uxsa.conf..411K}.

\begin{figure}[!h]
\includegraphics[width=80mm]{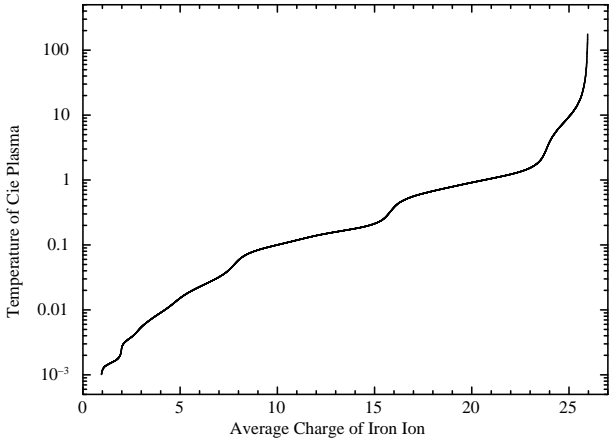}
 \caption{Relation between the average charge of iron ion 
 and the ionization temperature 
 calculated by the SPEX.}
   \label{fig:charge_plasmaT}
\end{figure}

In calculating 
the 
partial X-ray spectra,
we need to take 
into account the non-equilibrium between ions and electrons for part of the PSAC
and hence used the following two X-ray spectrum models; one is the
Cie (Collisional ionization equilibrium) model 
and the other is Neij (Non-Equilibrium Ionization Jump model;  \citet{1993A&AS...97..873K})
model, both included 
in the SPEX package. 
In general, equipartition between ions and electrons and ionization state of the ions in plasma reach 
thermal equilibrium if the 
product of the electron number 
density and the elapsed time since the shock becomes greater than 
$n_{\rm e}t  > 10^{12}$~cm$^{-3}$~s \citep{1984Ap&SS..98..367M}.
Accordingly, for the PSAC segments at which
$n_{\rm e}t  > 10^{12}$~cm$^{-3}$,
the Cie model is adopted
where 
the input temperature is common among 
the ions, the electrons and the ionization.
For the PSAC segments at which
$n_{\rm e}t  < 10^{12}$~cm$^{-3}$, on the other hand, the Neij model is adopted,
which is characterized by the 
three parameters 
$nt$, $T_{\rm ion}$ and $T$.
The $T_{\rm ion}$ and $T$ are initial ionization temperature 
and initial common temperature of the 
ions and the electrons
, respectively.
Thus, 
the Neij model can only 
treat the case that the ions and the electrons
share a common kinematical temperature, whereas our hydrodynamical model predicts that they in real have different temperatures near the top of the PSAC.
However, since the ionization process is governed by electron impacts to the ions, 
we adopt the electron temperature for $T$. 
$T_{\rm ion}$ 
is used as the ionization temperature that is common among all elements
reaching 
the start points of each segment.
The initial ionization temperature $T_{\rm ion}$ of a segment is evaluated from 
the average charge of iron ion achieved 
by the previous PSAC segment through 
a relation 
between the average charge of iron ion and the ionization temperature, 
which is 
calculated by the SPEX (figure \ref{fig:charge_plasmaT}). 
Only for the first segment 
laying on the top of the PSAC, $T_{\rm ion}$ is set to 0.002~keV, the limit of the Neij model.
%

\begin{figure}[!h]
\includegraphics[width=80mm]{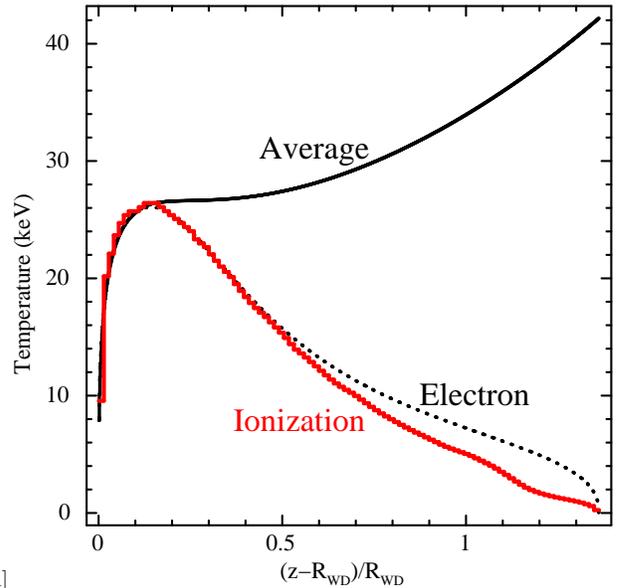}
 \caption{Averaged (black solid), electron (black dotted) and ionization (red solid) temperatures
 for the case $M_{\rm WD}$ = 1.2~$\Mo$ and $a = 0.001$~g~cm$^{-2}$~s$^{-1}$.}
   \label{fig:T_for_spe}
\end{figure}

Figure \ref{fig:T_for_spe} shows the ionization temperature calculated by the method defined above.
For this figure $\Mwd=1.2~\Mo$ and $a = 0.001$~g~cm$^{-2}$~s$^{-1}$ are adopted
where the non-equipartition are prominent 
(figure~\ref{fig:temperature_geocomp}).
The 
ionization temperature does not catch up 
that of electron in about 70\% of the PSAC from its top. 
We note, however, that the density of the segments in the ionization non-equilibrium area 
is smaller 
than those in equilibrium 
by 
one or two orders magnitudes,
which implies that the ionization non-equilibrium 
does not affect the resultant total X-ray spectrum significantly. 

\subsection{
X-ray spectra}

\begin{figure*}[!h]
\centering
\begin{tabular}{cc}
\includegraphics[width=80mm]{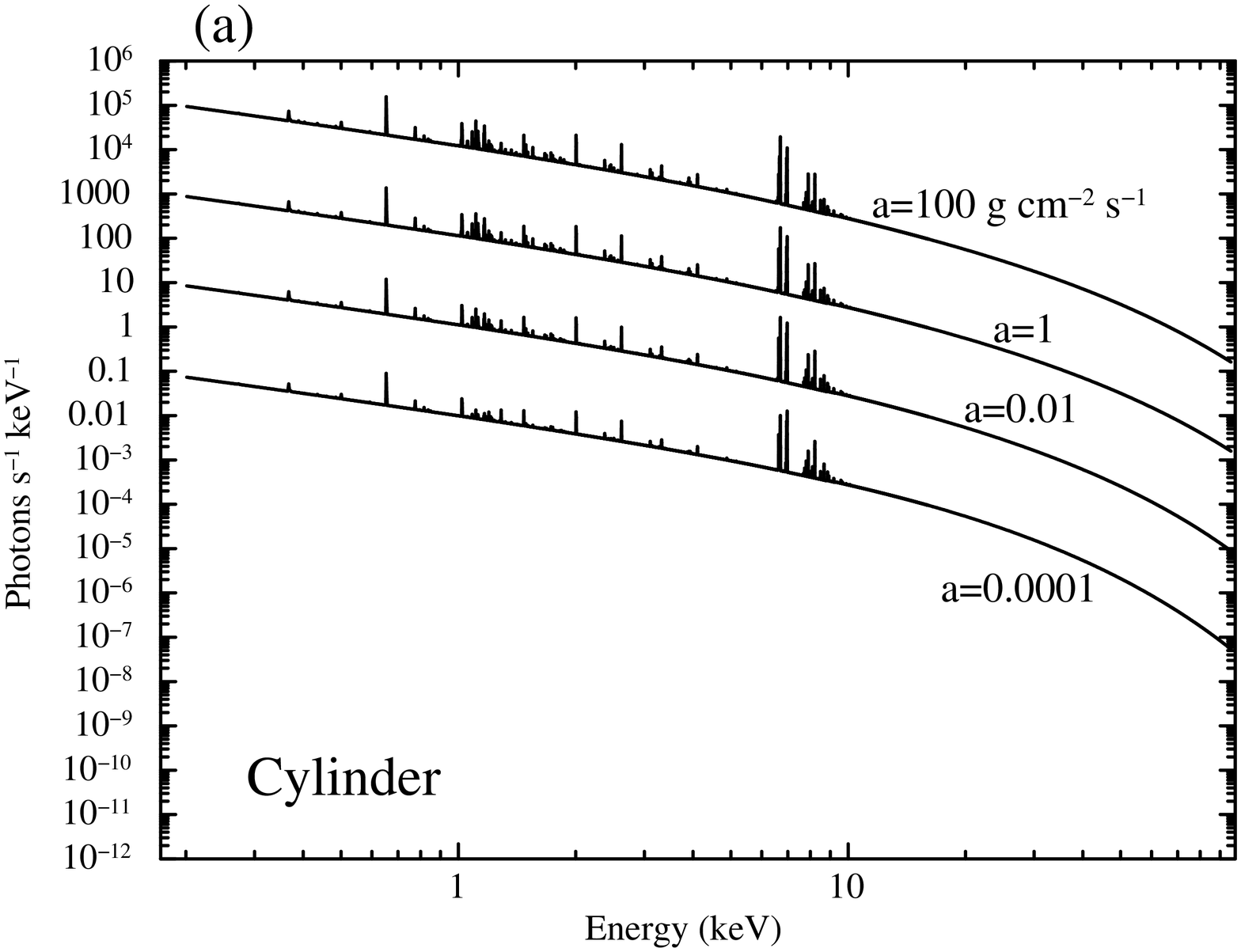}&
\includegraphics[width=80mm]{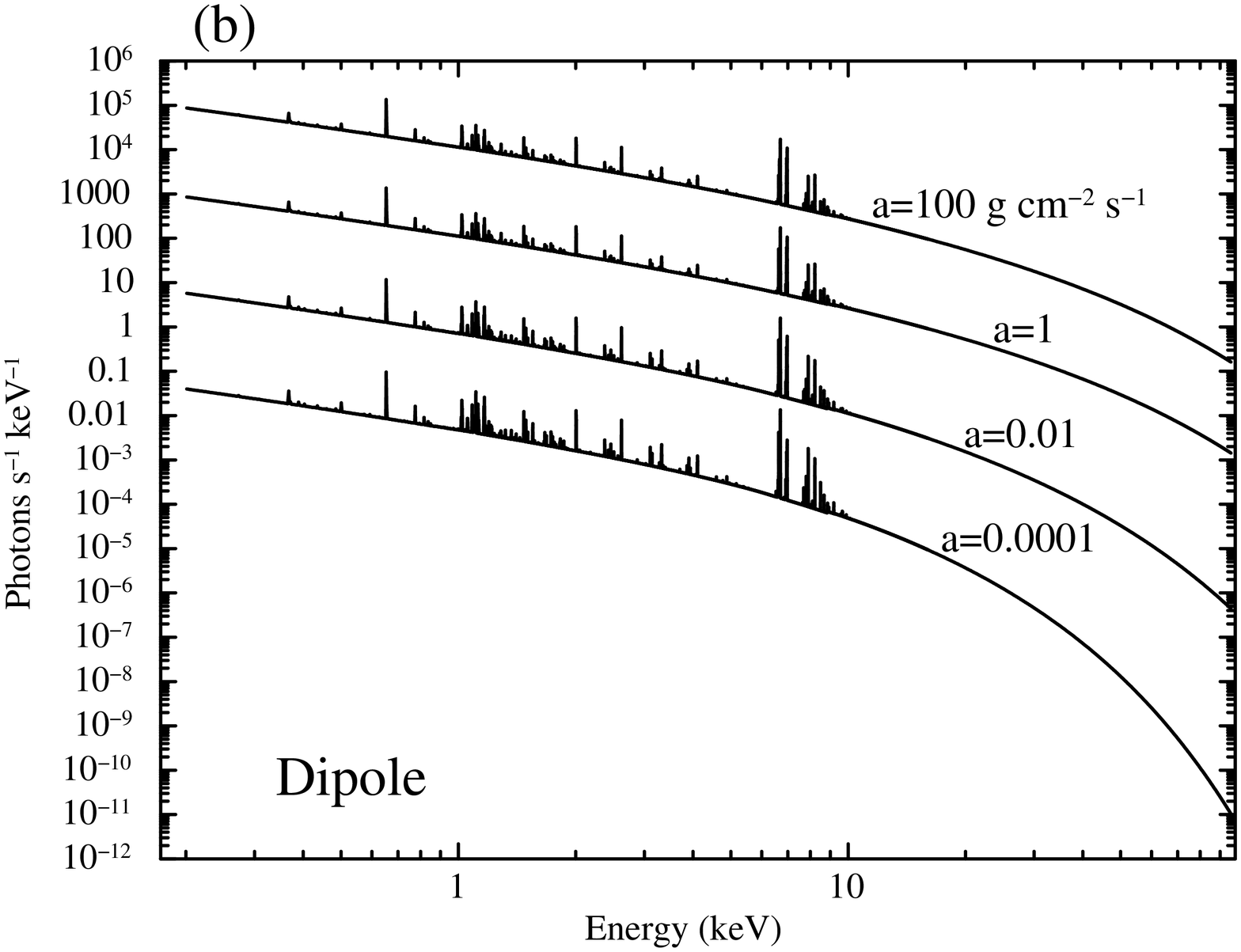}\\
\includegraphics[width=80mm]{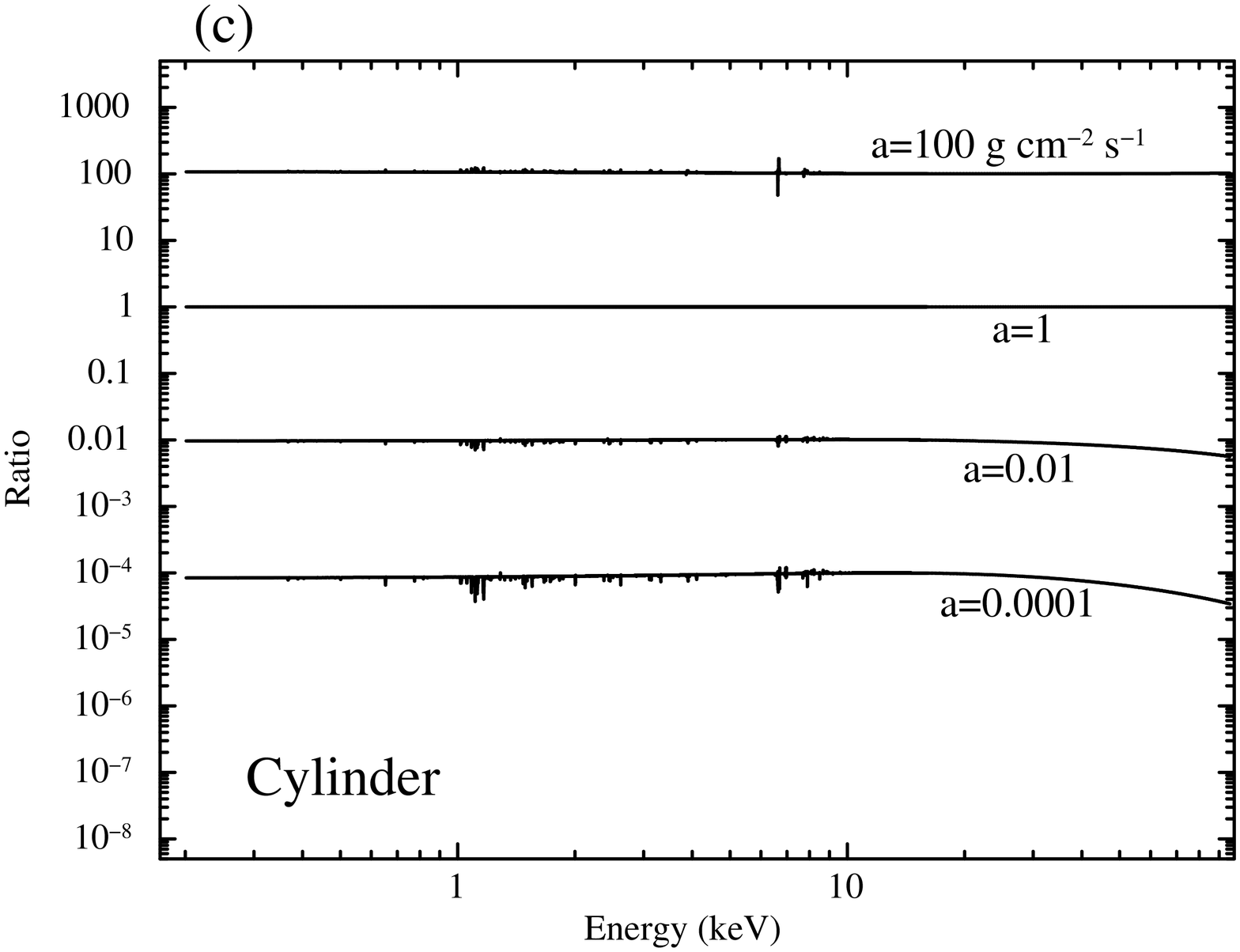}&
\includegraphics[width=80mm]{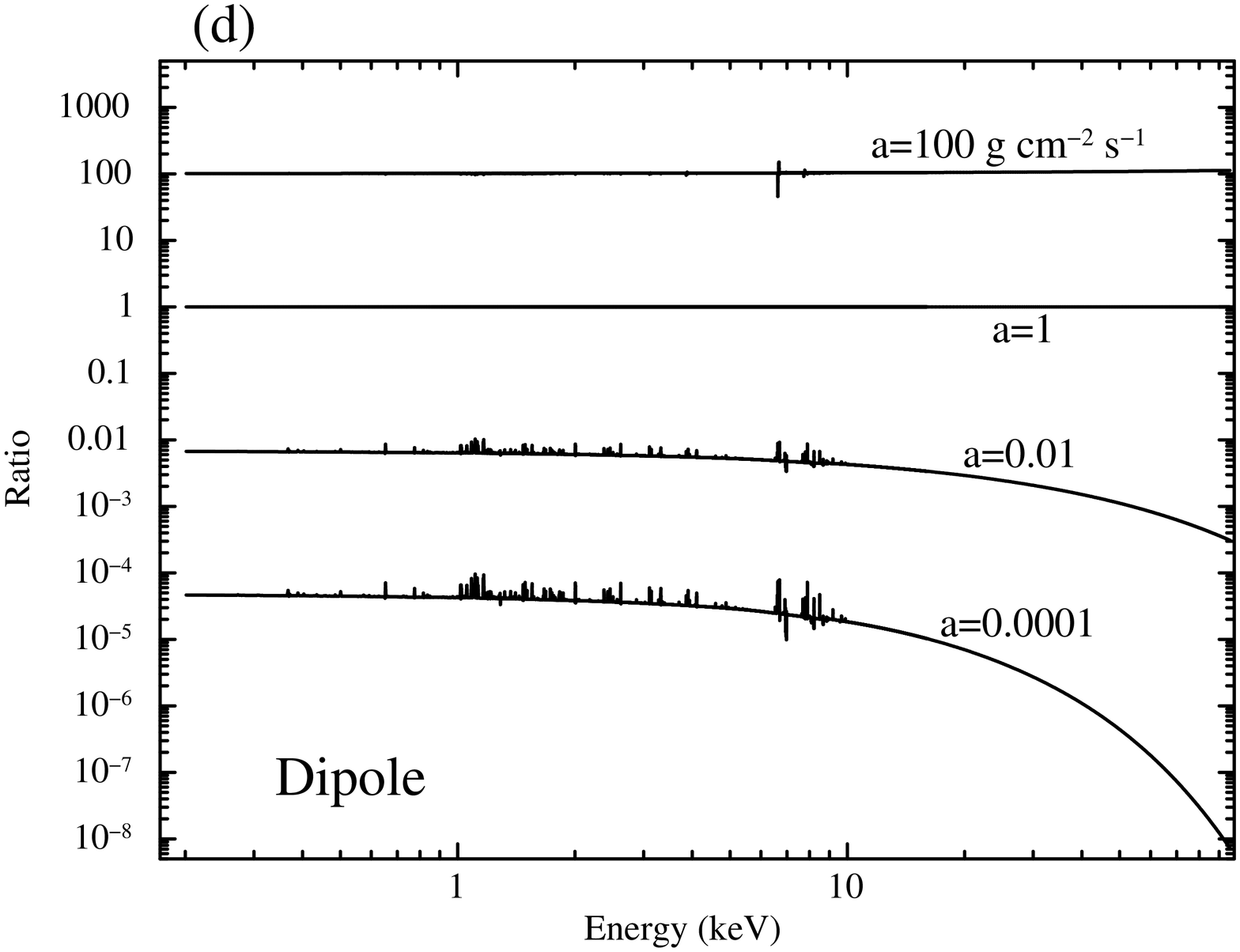}\\
\end{tabular}
\caption{X-ray spectra emitted from (a) cylindrical and (b) dipolar PSACs for $\Mwd = 0.7~\Mo$, 
and $a$ = 100, 1, 0.01 and 0.0001~g$^{-1}$~cm$^{-2}$~s$^{-1}$.
The spectral ratios to the spectrum with 
$a$ = 1~g$^{-1}$~cm$^{-2}$~s$^{-1}$ for the (c) cylindrical and (d) dipolar.
Abundance is assumed to be one solar. 
}
   \label{fig:spe}
\end{figure*}

\begin{figure}[!h]
\centering
\includegraphics[width=85mm]{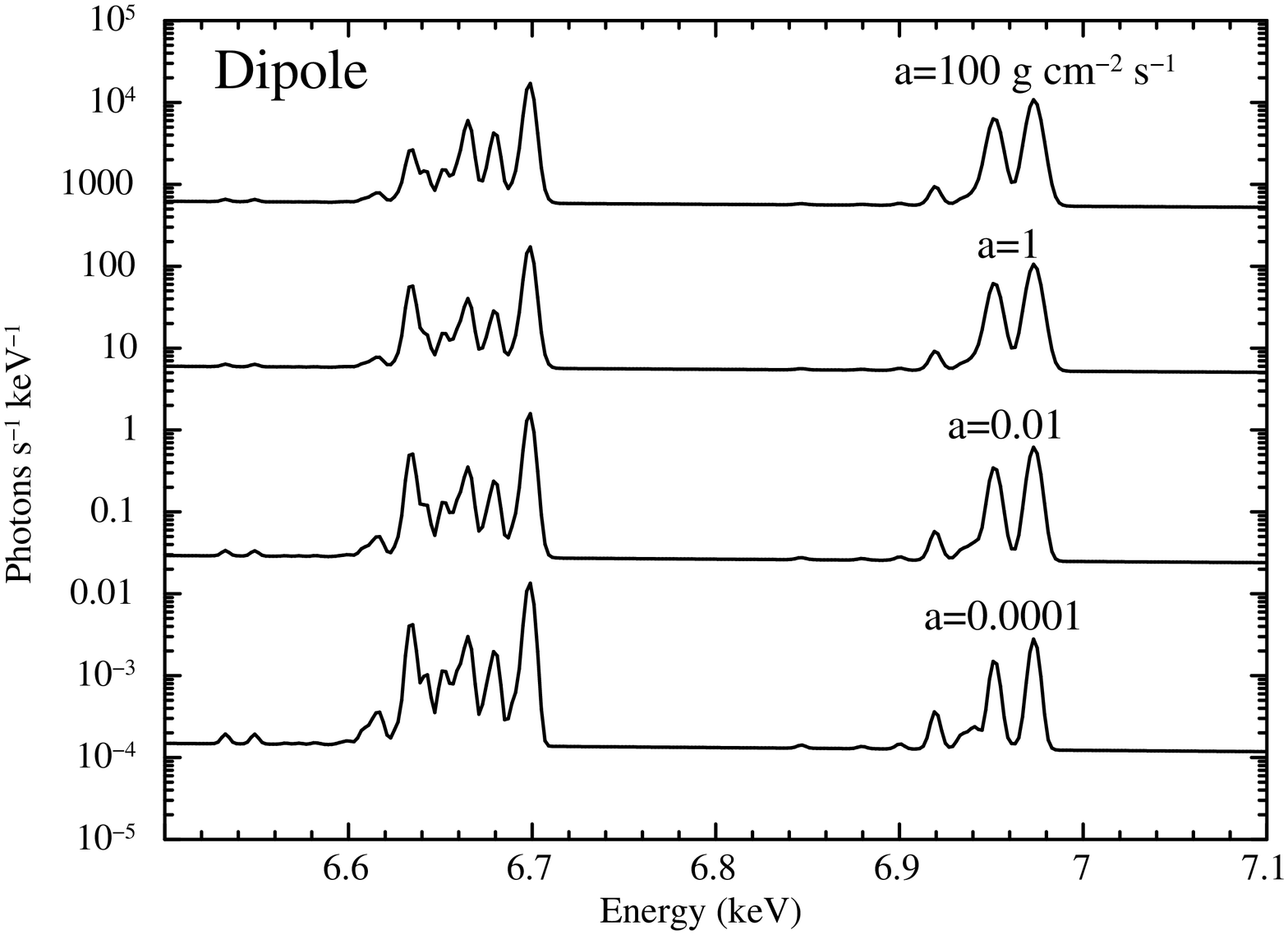}
\includegraphics[width=85mm]{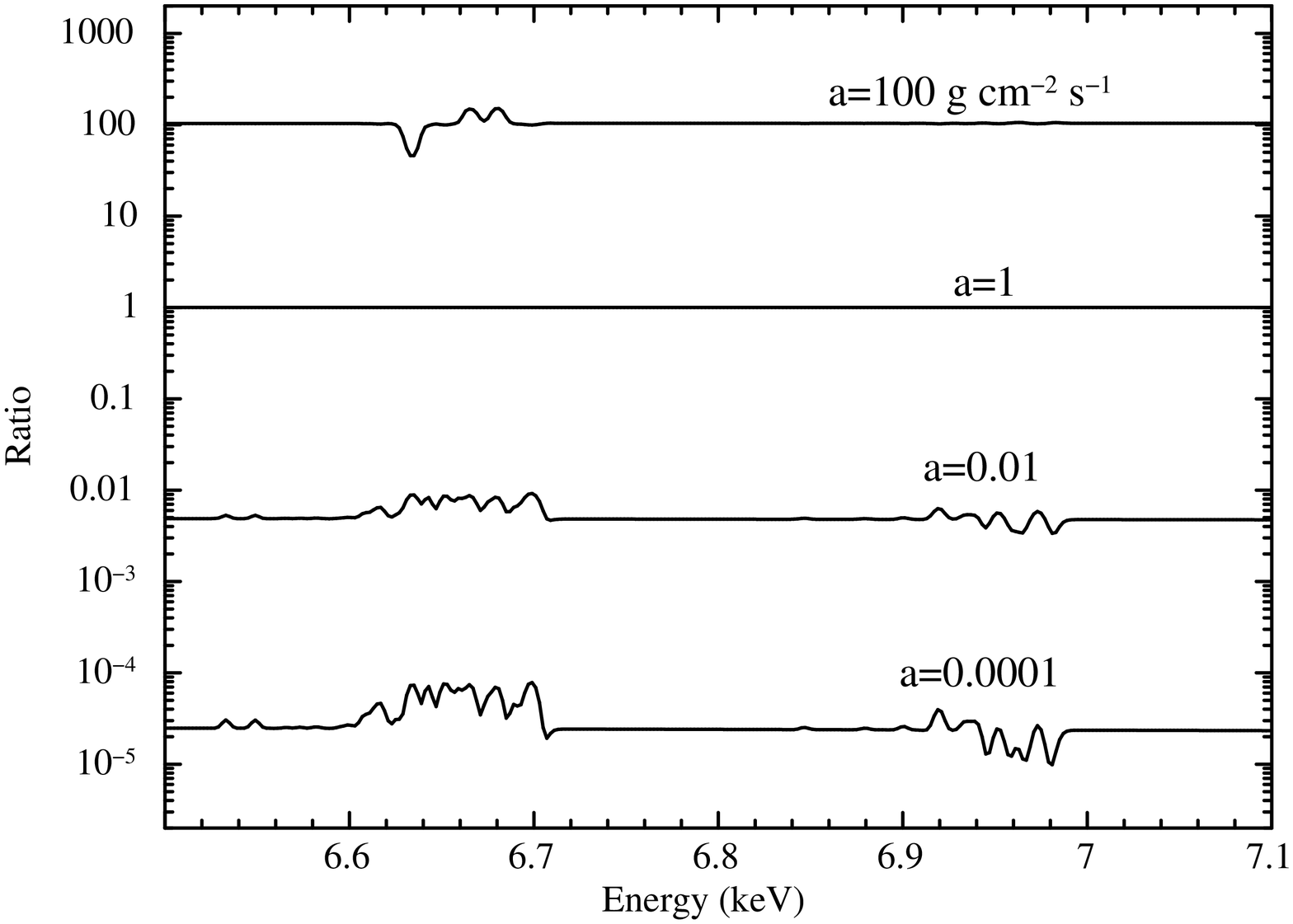}
\caption{Iron K$\alpha$ lines (top panel) emitted from dipolar PSACs for $\Mwd = 0.7~\Mo$, 
and $a$ = 100, 1, 0.01 and 0.0001~g$^{-1}$~cm$^{-2}$~s$^{-1}$.
The ratio to the iron K$\alpha$ lines with
$a$ = 1~g$^{-1}$~cm$^{-2}$~s$^{-1}$ (bottom panel).
Abundance is assumed to be one solar.}
   \label{fig:spe_iron}
\end{figure}

In figure \ref{fig:spe}, resultant spectra are shown for the cases 
$\Mwd = 0.7~\Mo$, $Z = Z_\odot$, 
$a$ = 100, 1, 0.01 and 0.0001~g~cm$^{-2}$~s$^{-1}$, 
as well as ratios of 
them to that of $a$ = 1~g~cm$^{-2}$~s$^{-1}$. 
The results are shown both for 
the cylinder 
and dipole geometries. 
In a higher 
specific accretion rate 
domain, 
the flux increases in proportion to the specific accretion rate.
Since the profiles 
of the temperature and density distributions 
resemble each other between the two PSAC geometries in the cases 
of $a$ = 1 and 100~g~cm$^{-2}$~s$^{-1}$, 
their spectral shapes are almost identical.
However, at He-like iron K$\alpha$ line ($\sim$6.7~keV), the spectral shapes of
the $a$ = 1 and 100~g~cm$^{-2}$~s$^{-1}$ cases are considerably different in both geometries.
This is 
because the density of PSAC reaches the critical density of iron $\sim 10^{18}$~cm$^{-3}$
between $a$ = 1 and 100~g~cm$^{-2}$~s$^{-1}$ 
and relative intensities of the He-like K$\alpha$ triplet alter with the density (figure \ref{fig:spe_iron}). 
On the other hand, the spectral shapes in the cases of 
$a$ = 0.01 and 0.0001~g~cm$^{-2}$~s$^{-1}$
clearly deviate 
from that of $a$ = 1~g~cm$^{-2}$~s$^{-1}$ for both PSAC geometries,
because the temperature of the PSACs are significantly reduced 
as the specific accretion rate becomes lower (figure \ref{fig:temperature_geocomp} and \ref{fig:a-Tmax}).
Change of ratios of H- and He-like iron K$\alpha$ lines (figure \ref{fig:spe_iron})
and deformation of the continuum above 10~keV energy bands are especially prominent,
and observationally important because the energy bands where these two phenomena emerge 
are hard to be influenced by 
complex and heavy absorbers 
generally detected in the X-ray spectra of IPs.
These spectral changes with
the specific accretion rate is larger for the dipolar PSAC, and
the spectrum of the dipolar PSAC is softer than that of the cylindrical 
at the same specific accretion rate. 
Since these deformations emerge 
in lower specific accretion rate systems,
the specific accretion rate should be considered as an important parameter 
in order to extract physical parameters such as WD masses from X-ray spectra.
At the same time, 
the spectral deformation 
potentially enables us
to measure the specific accretion rate with X-ray spectroscopy
which may give geometrical information of the height of PSAC,
the accreting area on the WD surface and shape of the PSAC.

Note that influence of the ionization non-equilibrium 
on X-ray spectrum is not significant
because of the low density of such PSAC domains 
for both geometries, 
although that effect is properly considered 
in this work.
For example, we obtained a result that the dipolar
 PSAC 
 with $\Mwd = 0.7~\Mo$, $a$ = 0.0001~g~cm$^{-2}$~s$^{-1}$ and $Z=Z_\odot$ 
at which
parameters the non-equilibrium effect manifests itself prominently, shows a 
spectrum 
more intense by about 4\% around 0.7~keV
and less intense by 0.01--0.02\% between 10 and 100~keV at most 
than 
the ionization 
equilibrium spectrum. 
At the energy around 0.7~keV, an iron L line forest 
is enhanced due to the ionization non-equilibrium effect because the ionization does not proceed 
compared with the equilibrium case. 
On the other band, bremsstrahlung radiation dominating over the latter energy band ($>10$~keV)
is weaker 
for the non-equilibrium model 
because of low electron 
temperature.
Even, 
in the 
energy band above 
5~keV that is 
important for the study of the PSAC for most mCVs \citep{1999ApJS..120..277E},
the difference is only about 0.2\%, which does 
not matter for the current observation quality.

The 
problem of the discrepancy of the WD mass measurement
(\S1) may be resolved if we treat the specific accretion rate as a free model parameter.
The X-ray spectrum of the PSAC is 
softer than that of 
the standard model
in the domain $a < a_{\rm crit}$,
which makes the resultant 
WD mass more massive.
We will 
verify this possibility by applying  
our model to observations in the forthcoming paper.

\section{Summary}

We calculated 
the density and temperature distributions
of the PSAC in IPs by 
 taking into account dependence on 
the specific accretion rate, the dipolar magnetic 
geometry, 
non-equipartition between electrons and ions, ionization non-equilibrium, 
and release of the gravitational potential with the form proportional to $r^{-1}$.
In particular, 
the specific accretion rate is floated 
over the wide range between 0.0001 to 100~g~cm$^{-2}$~s$^{-1}$.
This is the first comprehensive PSAC model that all these factors are fully taken into account.
With the density and temperature profiles, we constructed a
spectral model 
by dividing the PSAC radially into a hundred segments, 
and by integrating the spectrum from each segment calculated with 
the SPEX package. In our modeling, the free parameters are
the WD mass, the specific accretion rate and 
the metal abundance.

We found difference of the 
specific accretion rate significantly alters 
the profiles 
of the density and temperature distributions.
As long as the specific accretion rate is high enough, 
the density and temperature distributions of our modeling are 
consistent with those of the present standard model. 
There is, however, a critical specific accretion rate 
below which the profiles of the density and temperature distributions significantly deviate 
from those of the standard model. 
The standard model is no longer valid, if the specific accretion rate is below the critical value $a_{\rm crit}$, which is about 1 and 30 g~cm$^{-2}$~s$^{-1}$ for the 0.7 and 1.2$M_\odot$ WD, respectively, or when the heigh of the PSAC reaches about 1\% of the WD radius.
In addition, the profiles become different between the cylindrical and the dipolar geometries. 
The temperature profile shows the peak in the middle of the PSAC for the cylindrical geometry 
whereas that for the dipole geometry declines toward the WD surface more rapidly 
than that of the standard model.
Since the WD radius reduces for more massive WD,
the critical specific accretion rate 
is lower for the IPs holding the less massive WD.

The 
profiles 
of the density and 
temperature 
distributions
significantly changed with the
decrease of the specific accretion rate 
below the critical value, and mutual difference of the profiles between the two geometries is also enhanced. 
The non-equipartition between electrons and ions are significant in lower density domains, and hence, 
the non-equipartition is more significant for the dipolar PSAC.
With the low specific accretion rate of 0.0001~g~cm$^{-2}$~s$^{-1}$
the non-equipartition domain occupies 
80\% of the PSAC from its top for IPs holding a 0.7~$\Mo$ WD.
We also investigated the influence of the metal abundance, 
and found that it 
hardly influences the PSAC structure
between 0.1 and 2 
solar abundance which covers 
the entire IP population so far observed.

We calculated the X-ray spectra with the density and temperature distributions 
and 
found that the X-ray spectra depend on the specific accretion rate.
The X-ray spectra with the higher specific accretion rate than the critical value 
is almost constant 
except for the 
He-like iron K$\alpha$ emission line
due to the density dependence of the He-like triplet of the iron K$\alpha$ line.
When the specific accretion rate is smaller than 
the critical value, 
since the temperature in the PSAC reduces, the X-ray spectra soften,
which is more prominent for the dipolar PSAC. 
Specifically, the continuum component above 10~keV
and the ratio of H-like line to that of He-like  reduces as the specific accretion rate decreases.
The ionization non-equilibrium is not significant on X-ray spectra 
because of lower density than in equilibrium domain by a few orders of magnitudes.


\section*{Acknowledgement}
The authors would like to thank Prof. Ohashi T., Prof. Masai K. and Associate Prof. Ishisaki T.
for their very useful comments.


\begin{thebibliography}{99}
\bibitem[\protect\citeauthoryear{Aizu}{1973}]{1973PThPh..49.1184A} Aizu K., 1973, PThPh, 49, 1184 
\bibitem[\protect\citeauthoryear{Allan, Hellier, \& Beardmore}{1998}]{1998MNRAS.295..167A} Allan A., Hellier C., Beardmore A., 1998, MNRAS, 295, 167 
\bibitem[\protect\citeauthoryear{Beuermann \& Reinsch}{2008}]{2008A&A...480..199B} Beuermann K., Reinsch K., 2008, A\&A, 480, 199 
\bibitem[\protect\citeauthoryear{Brunschweiger et al.}{2009}]{2009A&A...496..121B} Brunschweiger J., Greiner J., Ajello M., Osborne J., 2009, A\&A, 496, 121 
\bibitem[\protect\citeauthoryear{Canalle et al.}{2005}]{2005A&A...440..185C} Canalle J.~B.~G., Saxton C.~J., Wu K., Cropper M., Ramsay G., 2005, A\&A, 440, 185 
\bibitem[\protect\citeauthoryear{Casares et al.}{1996}]{1996MNRAS.282..182C} Casares J., Mouchet M., Martinez-Pais I.~G., Harlaftis E.~T., 1996, MNRAS, 282, 182
\bibitem[\protect\citeauthoryear{Cropper, Ramsay, \& Wu}{1998}]{1998MNRAS.293..222C} Cropper M., Ramsay G., Wu K., 1998, MNRAS, 293, 222 
\bibitem[\protect\citeauthoryear{Cropper et al.}{1999}]{1999MNRAS.306..684C} Cropper M., Wu K., Ramsay G., Kocabiyik A., 1999, MNRAS, 306, 684
\bibitem[\protect\citeauthoryear{Eracleous, Halpern, \& Patterson}{1991}]{1991ApJ...382..290E} Eracleous M., Halpern J., Patterson J., 1991, ApJ, 382, 290 
\bibitem[\protect\citeauthoryear{Ezuka \& Ishida}{1999}]{1999ApJS..120..277E} Ezuka H., Ishida M., 1999, ApJS, 120, 277 
\bibitem[\protect\citeauthoryear{Frank, King, \& Raine}{1992}]{1992apa..book.....F} Frank J., King A., Raine D., 1992, apa..book,  
\bibitem[\protect\citeauthoryear{Fujimoto \& Ishida}{1997}]{1997ApJ...474..774F} Fujimoto R., Ishida M., 1997, ApJ, 474, 774 
\bibitem[\protect\citeauthoryear{Hachisu \& Kato}{2007}]{2007ApJ...662..552H} Hachisu I., Kato M., 2007, ApJ, 662, 552 
\bibitem[\protect\citeauthoryear{Hayashi et al.}{2011}]{2011PASJ...63S.739H} Hayashi T., Ishida M., Terada Y., Bamba A., Shionome T., 2011, PASJ, 63, 739 
\bibitem[\protect\citeauthoryear{H{\= o}shi}{1973}]{1973PThPh..49..776H} H{\= o}shi R., 1973, PThPh, 49, 776 
\bibitem[\protect\citeauthoryear{Imamura \& Durisen}{1983}]{1983ApJ...268..291I} Imamura J.~N., Durisen R.~H., 1983, ApJ, 268, 291 
\bibitem[\protect\citeauthoryear{Itoh et al.}{2006}]{2006ApJ...639..397I} Itoh K., Okada S., Ishida M., Kunieda H., 2006, ApJ, 639, 397 
\bibitem[\protect\citeauthoryear{Kaastra \& Jansen}{1993}]{1993A&AS...97..873K} Kaastra J.~S., Jansen F.~A., 1993, A\&AS, 97, 873 
\bibitem[Kaastra et al.(1996)]{1996uxsa.conf..411K} Kaastra, J.~S., Mewe, R., \& Nieuwenhuijzen, H.\ 1996, UV and X-ray Spectroscopy of Astrophysical and Laboratory Plasmas, 411 
\bibitem[\protect\citeauthoryear{Landi et al.}{2009}]{2009MNRAS.392..630L} Landi R., Bassani L., Dean A.~J., Bird A.~J., Fiocchi M., Bazzano A., Nousek J.~A., Osborne J.~P., 2009, MNRAS, 392, 630 
\bibitem[\protect\citeauthoryear{Masai}{1984}]{1984Ap&SS..98..367M} Masai K., 1984, Ap\&SS, 98, 367 
\bibitem[\protect\citeauthoryear{Nauenberg}{1972}]{1972ApJ...175..417N} Nauenberg M., 1972, ApJ, 175, 417 
\bibitem[\protect\citeauthoryear{Patterson}{1979}]{1979ApJ...234..978P} Patterson J., 1979, ApJ, 234, 978 
\bibitem[\protect\citeauthoryear{Ramsay et al.}{2000}]{2000MNRAS.316..225R} Ramsay G., Potter S., Cropper M., Buckley D.~A.~H., Harrop-Allin M.~K., 2000, MNRAS, 316, 225 
\bibitem[\protect\citeauthoryear{Saxton et al.}{2005}]{2005MNRAS.360.1091S} Saxton C.~J., Wu K., Cropper M., Ramsay G., 2005, MNRAS, 360, 1091 
\bibitem[\protect\citeauthoryear{Saxton et al.}{2007}]{2007MNRAS.379..779S} Saxton C.~J., Wu K., Canalle J.~B.~G., Cropper M., Ramsay G., 2007, MNRAS, 379, 779 
\bibitem[\protect\citeauthoryear{Schure et al.}{2009}]{2009A&A...508..751S} Schure K.~M., Kosenko D., Kaastra J.~S., Keppens R., Vink J., 2009, A\&A, 508, 751 
\bibitem[\protect\citeauthoryear{Spitzer}{1962}]{1962pfig.book.....S} Spitzer L., 1962, pfig.book
\bibitem[\protect\citeauthoryear{Suleimanov, Revnivtsev, \& Ritter}{2005}]{2005A&A...435..191S} Suleimanov V., Revnivtsev M., Ritter H., 2005, A\&A, 435, 191 
\bibitem[\protect\citeauthoryear{Woelk \& Beuermann}{1996}]{1996A&A...306..232W} Woelk U., Beuermann K., 1996, A\&A, 306, 232 
\bibitem[\protect\citeauthoryear{Wong \& Sarazin}{2009}]{2009ApJ...707.1141W} Wong K.-W., Sarazin C.~L., 2009, ApJ, 707, 1141 
\bibitem[\protect\citeauthoryear{Wu, Chanmugam, \& Shaviv}{1994}]{1994ApJ...426..664W} Wu K., Chanmugam G., Shaviv G., 1994, ApJ, 426, 664 
\bibitem[\protect\citeauthoryear{Wynn, King, \& Horne}{1997}]{1997MNRAS.286..436W} Wynn G.~A., King A.~R., Horne K., 1997, MNRAS, 286, 436 
\bibitem[\protect\citeauthoryear{Yuasa et al.}{2010}]{2010A&A...520A..25Y} Yuasa T., Nakazawa K., Makishima K., Saitou K., Ishida M., Ebisawa K., Mori H., Yamada S., 2010, A\&A, 520, A25 
\end{thebibliography}
\end{document}